\documentclass{article}
\usepackage{hyphenat}

\usepackage{PRIMEarxiv}

\usepackage[utf8]{inputenc} 
\usepackage[T1]{fontenc}    
\usepackage{hyperref}       
\usepackage{url}            
\usepackage{booktabs}       
\usepackage{amsfonts}       
\usepackage{nicefrac}       
\usepackage{microtype}      
\usepackage{lipsum}
\usepackage{fancyhdr}       
\usepackage{graphicx}       
\graphicspath{{media/}}     
\usepackage{subcaption}
\usepackage[authoryear,longnamesfirst]{natbib}

\title{Building AI Companions that Prioritise \\ Learning over Performance}

\author{
  Hassan Khosravi$^*$ \\
  The University of Queensland \\
  St Lucia, QLD 4072, Australia \\
  \texttt{h.khosravi@uq.edu.au} \\
  \And
  Dragan Ga\v{s}evi\'{c} \\
  Monash University \\
  Clayton, VIC 3800, Australia \\
  \texttt{dragan.gasevic@monash.edu} \\
  \And
  Shazia Sadiq \\
  The University of Queensland \\
  St Lucia, QLD 4072, Australia \\
  \texttt{shazia@eecs.uq.edu.au} \\
  \And
  Lixiang Yan \\
  Tsinghua University \\
  Haidian District, Beijing 100084, China \\
  \texttt{lixiangyan@tsinghua.edu.cn} \\
  \And
  Jason M. Lodge \\
  The University of Queensland \\
  St Lucia, QLD 4072, Australia \\
  \texttt{jason.lodge@uq.edu.au} \\
  \And
  Jason M. Tangen \\
  The University of Queensland \\
  St Lucia, QLD 4072, Australia \\
  \texttt{j.tangen@uq.edu.au} \\
  \And
  Paul Denny \\
  The University of Auckland \\
  Auckland 1010, New Zealand \\
  \texttt{p.denny@auckland.ac.nz} \\
  \And
  Kristen DiCerbo \\
  Khan Academy \\
  Mountain View, CA 94041, USA \\
  \texttt{kristen@khanacademy.org} \\
  \And
  Simon Buckingham Shum \\
  University of Technology Sydney \\
  Ultimo, NSW 2007, Australia \\
  \texttt{simon.buckinghamshum@uts.edu.au} \\
  \And
  Ryan S. Baker \\
  Adelaide University \\
  Adelaide, SA 5005, Australia \\
  \texttt{ryan.baker@adelaide.edu.au} \\
}

\begin{document}
\maketitle

\begin{abstract}
Large language models (LLMs) are rapidly transforming knowledge work by improving the quality and efficiency of tasks such as writing, coding, and data analysis. However, their growing use in education has exposed a learning–performance paradox: while they can enhance short-term task performance, they may also undermine genuine learning, including cognitive growth, knowledge transfer, and metacognitive development. This paper addresses the question of how artificial intelligence should be designed and used to support learning rather than merely improve immediate outputs. We introduce the concept of AI learning companions, defined as adaptive, pedagogically informed, LLM-powered agents designed for integration into learning environments.  We propose a framework for their design built on three interrelated foundations: a pedagogical foundation focused on how students learn with AI, an adaptive foundation focused on how AI learns about students, and a responsible design foundation ensuring systems remain transparent, accountable, inclusive, and secure. The framework is illustrated through five case studies spanning diverse educational contexts, levels, and tool designs, revealing both the promise and current limitations of existing tools. We conclude that there is a necessary shift away from LLMs designed for task-oriented performance, and beyond simply prompting them to act as tutors, toward deliberately developed AI learning companions that are pedagogically sound, adapt to their learners, and foster durable understanding, metacognitive growth, and learner agency.
\end{abstract}


\section{Introduction}
Generative artificial intelligence (AI), particularly large language models (LLMs), is rapidly transforming the nature of knowledge work by enhancing productivity and reshaping professional practices. Unlike earlier forms of automation that targeted repetitive, low-cognition tasks, LLMs extend automation into domains requiring advanced cognitive skills such as writing, analysis, programming, and decision-making \cite{cazzaniga2024}. This marks a profound shift from automating routine labour to augmenting complex, knowledge-intensive work.

This shift aligns with the paradigm of co-intelligence or \emph{hybrid intelligence}, in which human and artificial intelligences are deliberately combined to exceed the capabilities of either alone \cite{jarvela2025}. AI systems contribute speed, scale, and precision to data-intensive processes, while humans provide contextual awareness, ethical reasoning, and creative synthesis \cite{dellermann2019}. A growing body of field studies substantiates these benefits, demonstrating substantial time savings and performance boosts across sectors. For instance, studies on knowledge workers using Microsoft 365 Copilot report significant reductions in time spent on email and document drafting \cite{dillon2025a}, while engineers using GitHub Copilot completed coding tasks 56\% faster \cite{peng2023}. This productivity leap is reinforced by experimental evidence in professional writing tasks, where generative AI was found to reduce task completion time by 40\% while simultaneously increasing output quality by 18\% \cite{noy2023experimental}. Similarly, AI augmentation has been shown to increase customer service agent efficiency by 14\% and help less experienced workers narrow the skills gap with experts by enabling newer employees to move up the productivity curve more quickly than traditional training allows \cite{brynjolfsson2025generative, dellacqua2023}.

LLMs are rapidly becoming embedded in students' everyday study practices, supporting tasks such as summarising, reasoning, writing, coding, and analysis \cite{chung2026}. As education systems shift from prohibition towards more structured adoption, many students remain uncertain about how to use AI effectively for learning, with some voicing concerns about cheating themselves of genuine understanding \cite{chung2026}. These concerns are consistent with UNESCO and OECD warnings about the widening gap between the pace of technological change and the pedagogical frameworks needed to guide its effective use in education \cite{unesco2023, oecd2026}. They also echo longstanding findings from the learning sciences that improvements in immediate task performance do not necessarily lead to durable learning \cite{soderstrom2015, yan2025}. While recent meta-analyses report that students using ChatGPT or similar systems show improved performance on assignments and exams \cite{deng2025, han2025}, these findings must be interpreted with caution: the studies on which they draw largely measure short-term, scaffolded performance rather than durable learning, and have been critiqued as an effect in search of a cause \cite{weidlich2025}.

The distinction between performance and learning is not merely semantic. It reflects what the literature terms the learning-performance paradox \cite{soderstrom2015, yan2025}: the well-documented phenomenon whereby AI tools can enhance short-term task outputs while simultaneously undermining durable learning, including cognitive growth, knowledge transfer, and metacognitive development. This paradox arises from a well-documented cognitive mechanism: cognitive offloading, defined as the use of physical action to alter the information processing requirements of a task so as to reduce cognitive demand \cite{risko2016}. When students rely on AI for sensemaking, planning, monitoring, and evaluation, they reduce engagement in the very processes that make self-regulated learning effective \cite{fan2024, zhai2024}. This matters because durable understanding emerges through grappling with complexity rather than bypassing it \cite{kapur2016}. The most compelling empirical evidence for this comes from a large randomised experiment in high school mathematics, in which students given access to an AI assistant showed improved problem-solving during learning but suffered significant harm to their durable, independent learning once the AI was removed \cite{bastani2025}. The same pattern has been observed in misconception correction: Corbett et al., \cite{corbett2026} found that personalised AI dialogue produced significantly larger immediate belief reductions than traditional textbook refutation, but by the two-month follow-up the two conditions had converged, suggesting that the engagement and confidence generated by AI interaction did not translate into durable learning advantages. AI-supported tasks can thus produce stronger artefacts without equivalent learning gains, alongside reduced self-regulatory monitoring \cite{darvishi2024}, increased dependence on AI assistance, and a form of metacognitive laziness in which learners abdicate the cognitive and metacognitive effort required for deep understanding \cite{fan2024}. Optimising for immediate task completion therefore comes at the cost of cultivating the processes that support durable learning \cite{yan2025, soderstrom2015}.

These findings illuminate a fundamental tension. While LLMs show clear potential for supporting hybrid intelligence and enhancing productivity in professional contexts, education requires a different orientation, one that values process over product, prioritises cognitive growth over immediate results \cite{bjork2013, soderstrom2015}, and fosters metacognitive development through reflection, self-regulation, and strategic awareness of one's own learning processes \cite{zimmerman2002}. This tension raises a central question: how should AI be designed and used to cultivate learners' cognitive and metacognitive abilities rather than merely support short-term performance?

In response, prompt-level approaches have emerged to steer LLMs towards more pedagogically productive interaction. Commercial systems now include study-oriented modes that use questioning, step-by-step guidance, and checks for understanding rather than simply returning answers. Although empirical studies indicate that prompt-based guardrails can mitigate many of the negative effects associated with direct, unstructured LLM use \cite{bastani2025}, their effects on learning are often near null, and they are unlikely to match the learning benefits historically associated with Intelligent Tutoring Systems (ITSs) \cite{borchers2025}. This is because prompt-based approaches remain limited in three important respects: first, they are typically reactive rather than proactive; second, they are largely stateless in relation to learners' prior knowledge, goals, and misconceptions; and third, they operate at the level of specific tasks rather than supporting broader learning trajectories across the curriculum. By contrast, ITSs and adaptive learning systems have long addressed these challenges through student modelling and knowledge tracing \cite{abdelrahman2023}, and through instructional adaptation engines that personalise feedback, scaffolding, and learning pathways \cite{aleven2016} with strong evidence of learning benefits \cite{xu2019}. Yet these systems have been difficult to scale because they rely heavily on manually authored content, feedback rules, and pedagogical pathways \cite{dermeval2018}. A significant gap therefore remains in how to integrate the conversational flexibility and scalability of LLMs with the learner modelling and pedagogical adaptivity of ITSs to create AI-powered educational tools that genuinely support learning.

Building on decades of research in the learning sciences, adaptive and intelligent tutoring systems, and human-centred approaches to AI, we introduce the concept of AI learning companions: adaptive, pedagogically informed, LLM-powered agents designed for integration into diverse educational environments and explicitly engineered to prioritise durable learning over short-term performance. We propose a framework for their design built on three interrelated foundations. The pedagogical foundation addresses how students learn with AI, fostering cognitive engagement, metacognitive awareness, and a sense of agency throughout the learning process. The adaptive foundation addresses how AI learns about students, modelling cognitive, affective, and behavioural data to tailor feedback, adapt instruction, and sustain personalised, data-informed feedback loops. The responsible design foundation ensures that AI companions remain transparent, inclusive, accountable, and secure, upholding human oversight, equity, and trust. In what follows, we first conceptualise learning companions by synthesising insights from the relevant literature to clarify their defining characteristics and affordances. We then elaborate on the design and implementation of such companions across the three foundations, detailing how each contributes to building AI-based learning companions, and illustrate the framework through five case studies spanning diverse educational contexts, levels, and tool designs.

\section{Conceptualising Learning Companions}
The preceding analysis of the learning-performance paradox points to a fundamental design problem: the LLMs now entering education were not built for education. They were built for work, and the logic of that design, optimising for output quality, minimising cognitive effort, and treating each interaction as independent and disposable, is precisely what makes them unsuitable as learning tools without deliberate redesign. Understanding what AI for learning must look like therefore requires first being clear about how it differs from AI for work, not just in context but in purpose, design logic, and what it means to succeed.

\begin{figure}[t]
    \centering
    \includegraphics[width=0.9\textwidth]{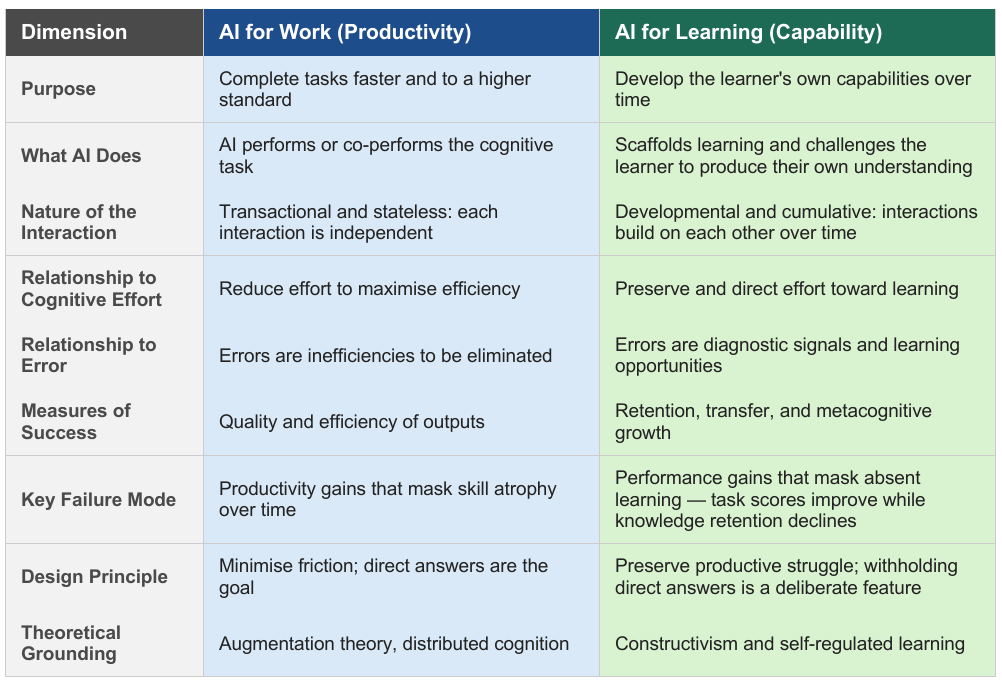}
    \caption{Comparison of AI designed for work and AI designed for learning, highlighting the reorientation required when moving from professional productivity contexts to educational environments.
    }
    \label{tab:comparison}
\end{figure}

Figure~\ref{tab:comparison} formalises this contrast across nine dimensions. The most critical distinctions concern purpose, the nature of the interaction, and the relationship to cognitive effort. In work contexts, AI performs or co-performs the cognitive task on behalf of the user, each interaction is transactional and stateless, and success is measured by the quality and efficiency of outputs. In learning contexts, AI must instead scaffold and challenge the learner to produce their own understanding, interactions must be developmental and cumulative, and success is measured by retention, transfer, and metacognitive growth. Equally revealing are the key failure modes: AI for work fails when productivity gains mask skill atrophy over time, while AI for learning fails when task scores improve while knowledge retention declines --- the learning-performance paradox made concrete \cite{bastani2025, weidlich2025}. These failure modes are not accidental but follow directly from design intent. AI for work is designed to minimise friction and provide direct answers; AI for learning must deliberately withhold direct answers and preserve productive struggle as the mechanism through which durable understanding is built \cite{bjork2013, soderstrom2015}. Together these dimensions reveal that the reorientation required is not a minor adjustment to prompting strategy or interface design. It demands a fundamental reconceptualisation of what AI systems in education are designed to do, how they interact with learners, and how their effectiveness is evaluated.

\begin{figure}[t]
    \centering
    \includegraphics[width=1\textwidth]{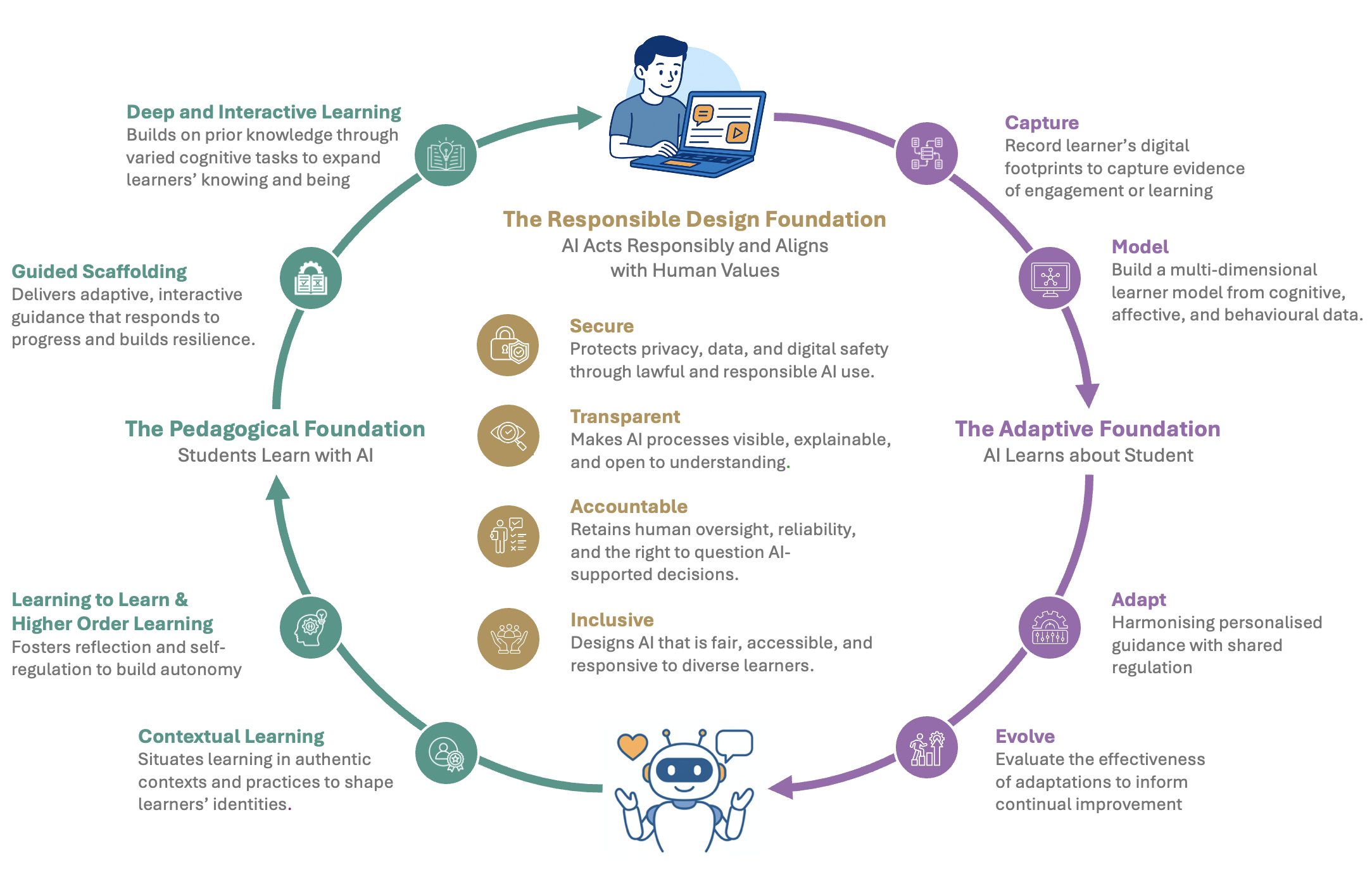}
    \caption{Framework for building AI companions to support learning. 
    The left cycle represents how \textit{students learn with AI} through deep learning, guided scaffolding, metacognitive growth, and contextual engagement. 
    The right cycle represents how \textit{AI learns about students} by capturing, modelling, adapting, and evaluating learning processes. 
    At the centre, \textit{ethical principles}—accountability, transparency, inclusivity, and security—govern the interactions between the two loops.}
    \label{fig:learning-loops}
\end{figure}

We propose AI learning companions as the design response to this reorientation: adaptive, pedagogically informed, LLM-powered agents designed for sustained integration into diverse educational environments and explicitly engineered to prioritise durable learning over short-term performance. Learning companions comprise three interrelated dimensions that together define how AI can meaningfully support learning as visualised in Figure \ref{fig:learning-loops}: (1) how students learn with AI, guided by pedagogical principles from the learning sciences; (2) how AI learns about students, through adaptive mechanisms that model and respond to individual progress; and (3) how ethical and human-centred principles ensure these systems remain trustworthy, equitable, and aligned with human values.

The \textbf{pedagogical foundation} addresses how students learn with AI. The challenge for AI design is not to reinvent pedagogy but to determine how generative systems can provide the precise support necessary to cultivate both students' cognitive abilities and their metacognitive capacities. To guide this design, we draw on the Higher Education Learning Framework (HELF) \cite{nugent2019}, which synthesises the learning sciences literature into a set of core pedagogical principles. Building on the HELF, we derive four principles for AI companion design: \textit{deep and interactive learning}, which prompts learners to actively process and connect new knowledge through varied cognitive tasks; \textit{guided scaffolding}, which delivers adaptive guidance calibrated to the learner's Zone of Proximal Development; \textit{learning to learn and higher-order learning}, which fosters metacognitive skills, reflection, and learner autonomy; and \textit{contextual learning}, which situates learning in authentic contexts to shape disciplinary identity and ensure knowledge is transferable. Each principle and its implications for companion design are elaborated in Section~\ref{sec:learn-with-ai}.

The \textbf{adaptive foundation} addresses how AI learns about students. While LLMs offer conversational flexibility and scalability that traditional intelligent tutoring systems could not match, most current LLM applications remain stateless --- each interaction begins without memory of the last, making sustained adaptation to individual learners impossible. This foundation resolves that limitation by organising adaptivity into a continuous four-stage cycle: \textit{Capture}, which records learners' digital footprints to gather evidence of engagement and understanding; \textit{Model}, which builds a multi-dimensional learner model from cognitive, affective, and behavioural data; \textit{Adapt}, which determines what, how, and when to adjust learning experiences; and \textit{Evolve}, which evaluates the effectiveness of those adaptations to drive continual improvement. Each stage and its implications for companion design are elaborated in Section~\ref{sec:ai-learns-about-student}.

The \textbf{responsible design foundation} addresses how AI companions can act with integrity and uphold human values. The highly personalised nature of adaptive AI demands that companions be not only pedagogically effective but also trustworthy, equitable, and aligned with educational values. To guide this design, we organise responsible practice around four interrelated commitments: \textit{security}, which protects learner privacy, data, and digital wellbeing through lawful and responsible data practices; \textit{transparency}, which makes AI processes, reasoning, and data flows visible and explainable to learners and educators; \textit{accountability}, which retains meaningful human oversight and preserves learners' rights to question AI-supported decisions; and \textit{inclusion}, which ensures companions are fair, accessible, and responsive to diverse learners, abilities, languages, and contexts. Each commitment and its implications for practice are elaborated in Section~\ref{sec:ethical-companions}.

\section{The Pedagogical Foundation: Students Learn with AI}\label{sec:learn-with-ai}
Early applications of LLMs in education have largely relied on prompting strategies that cast the model in roles such as a Socratic tutor (e.g., Khan Academy’s Khanmigo) or study modes offered by OpenAI’s ChatGPT and Google’s Gemini. While useful, these approaches often remain limited in scope and provide only surface-level support. To move beyond this, we draw on evidence from the learning sciences about how people learn best and employ the Higher Education Learning Framework (HELF) \cite{nugent2019} to propose the design of effective AI companions built around four interrelated principles that are essential for fostering robust and equitable learning. These principles are discussed in detail in the following sections.

\subsection{Deep and Interactive Learning.} 
A central goal of education, and one that AI companions must be explicitly designed to support, is deep learning: the kind of durable understanding that transfers across contexts and persists long after instruction ends. Cognitive science provides clear guidance on what produces it. The \textit{generation effect} shows that knowledge constructed by the learner is retained far better than knowledge passively received \cite{roediger2011}; the \textit{testing effect} demonstrates that retrieving information from memory strengthens it more than re-reading or re-watching \cite{roediger2006}; and research on \textit{desirable difficulties} establishes that conditions which feel effortful in the moment, including retrieval practice, interleaving, and spacing, produce stronger long-term learning than conditions that feel fluent \cite{bjork2011}. The practical implication is direct: what matters is not whether a companion presents information well, but whether it makes the student do the thinking. A companion that explains clearly but lets the learner remain passive is, from a learning science perspective, not much better than a well-produced video.
Learning is also inherently social. Teaching and learning are grounded in humans' innate capacity and motivation to connect with others \cite{immordino2007we}, and interaction is not peripheral but integral to meaningful learning. Social dynamics shape motivation, engagement, and higher-order thinking through processes including action understanding, empathy, and theory of mind \cite{klimecki2013empathy}, the same mechanisms that generate shared understanding and collective engagement in human learning environments \cite{wheatley2012mind}. For AI companions, this means that the quality of interaction matters as much as the quality of content: dialogue, perspective-taking, and collaborative sense-making are not add-ons but core mechanisms through which learners construct understanding.
\textit{Strategies for AI companions.} To embody these principles, AI companions should be designed to prompt generation and retrieval rather than provision. Rather than explaining a concept, a companion should ask the learner to recall it, apply it, or explain it in their own words. Rather than confirming correctness, it should invite elaboration, surface connections to prior knowledge, and expose gaps in understanding. Companions can also shift fluidly across interactional roles, as tutor, peer, or novice, to sustain dialogue and foster the kind of perspective-taking that amplifies interactive learning. The criterion for good design is not whether the companion feels helpful in the moment but whether it leaves the learner having done more thinking than they would have done alone.

\subsection{Guided Scaffolding} 
A central condition for meaningful learning is that it involves challenge. Cognitive science emphasises that learning is optimised when tasks are neither trivial nor overwhelming, but situated in a “zone of proximal development” where learners can stretch their current abilities with appropriate guidance \cite{vigotsky1978}. Research on desirable difficulties highlights that introducing challenges such as retrieval practice, interleaving, and spacing promotes deeper encoding and transfer, even though these conditions may feel effortful in the moment \cite{roediger2011,bjork2011}. Challenge, however, is not synonymous with frustration. Productive struggle occurs when difficulty is balanced with support, enabling learners to persist, adapt strategies, and ultimately experience mastery. Without this balance, excessive difficulty can lead to disengagement, whereas insufficient challenge fosters surface processing and complacency. Thus, challenge and difficulty should be understood as essential mechanisms through which learners refine strategies, consolidate understanding, and develop resilience.

Learning is not solely a cognitive process but is deeply interwoven with emotion. Neuroscience and educational psychology consistently show that emotions influence attention, memory, and motivation \cite{immordino2007we,tyng2017influence}. Positive emotions such as curiosity, enjoyment, and hope broaden learners’ willingness to explore and engage, while negative emotions like anxiety or boredom can constrain focus and impede progress. Importantly, negative emotions are not uniformly detrimental: when framed within a supportive environment, experiences of confusion, cognitive dissonance, or even frustration can catalyse deeper inquiry and problem solving \cite{d2014confusion}. Emotions also serve social and motivational functions, shaping how learners connect with others and how they interpret the value and relevance of their learning experiences. Understanding the role of affect underscores that robust learning environments must not only target knowledge and skills but also cultivate emotional conditions that sustain engagement, resilience, and growth.

\textit{Strategies for AI companions.} To embody these principles, AI companions should be designed to scaffold productive challenge while also attuning to learners’ emotional states. They can introduce manageable levels of difficulty such as prompting elaboration, posing counterexamples, or encouraging retrieval while offering adaptive support that prevents unproductive frustration. At the same time, AI companions can monitor affective cues in language and interaction, responding with encouragement, reframing confusion as a learning opportunity, or offering strategies for regulation. By balancing challenge with support and integrating sensitivity to emotion, AI companions can help learners experience difficulty as an engine of growth rather than a barrier. In doing so, they contribute to environments where cognitive stretch is paired with emotional scaffolding, enabling learners to persist, adapt, and flourish.

\subsection{Learning to Learn and higher order learning} 
A central condition for sustained educational growth is the capacity to learn how to learn. This involves developing metacognitive awareness: knowing what one knows, monitoring progress, and recognising when strategies need adjustment. Research on self-regulated learning highlights that learners who actively plan, monitor, and evaluate their approaches demonstrate greater autonomy, adaptability, and persistence \cite{zimmerman2002, pintrich2004role}. Such metacognitive processes are not innate; they must be deliberately cultivated through opportunities for reflection, feedback, and strategy use.

Metacognition is not merely a companion to higher-order thinking but its prerequisite. Meaningful education requires engagement with processes that extend beyond the reproduction of knowledge toward its transformation and application. Bloom's taxonomy and subsequent frameworks emphasise the importance of analysis, synthesis, evaluation, and creation as core dimensions of higher-order thinking \cite{bloom1956, anderson2001taxonomy}, and learners develop deeper understanding when they critique arguments, integrate knowledge across domains, and generate novel ideas. Higher-order learning is also closely linked to transfer, the ability to apply principles in new or unfamiliar settings \cite{perkins1993teaching, bransford2000}, which requires not only domain knowledge but also the metacognitive awareness to recognise when and how prior learning applies. Together, these capacities are vital for preparing students to navigate complex, real-world problems where adaptability, creativity, and critical judgment are as important as content mastery.

\textit{Strategies for AI companions.} To embody these principles, AI companions should be designed to scaffold both metacognitive awareness and higher-order thinking as interrelated capacities. They can prompt learners to articulate their goals, plan strategies, and reflect on progress, thereby strengthening self-regulation and adaptability. A particularly powerful application is metacognitive calibration: comparing a learner's expressed confidence against their actual performance to surface the gap between what they think they know and what they demonstrably know \cite{kornell2009}. Students are consistently poor judges of their own understanding, and AI companions are well positioned to make this visible, not by correcting learners directly, but by designing interactions that reveal the limits of their current knowledge and prompt them to adjust their self-assessment accordingly. At the same time, companions can challenge learners with open-ended questions, request justifications for reasoning, and encourage comparisons across contexts to deepen analysis and synthesis. By prompting learners to generate explanations or teach back concepts, AI companions make higher-order thinking processes explicit and accessible, helping learners develop the reflective and critical capacities required for transfer, innovation, and lifelong learning.

\subsection{Contextual Learning} 
\label{sec:contextual_learning}
Contextual learning emphasises that knowledge is not acquired in isolation but is shaped by the environment, tools, and practices in which it is used. Situated learning theory highlights that authentic contexts provide learners with opportunities to engage in tasks that mirror real-world practices, enabling knowledge to be both applied and understood in relation to its use \cite{brown1989}. Kolb’s experiential learning cycle highlights the iterative interplay of experience, reflection, conceptualisation, and experimentation \cite{kolb2014experiential}, while Shulman’s signature pedagogies and the theory of threshold concepts emphasise the transformative practices that shape professional identity \cite{shulman2005,meyer2003}.

Learning is also a process of becoming. From this perspective, education is not simply about acquiring skills or information but about shaping identities, dispositions, and ways of participating in communities of practice \cite{wenger1998}. Becoming a learner in a discipline involves appropriating its language, methods, and values while negotiating one’s own sense of purpose and belonging. Such identity work is developmental, as students shift from peripheral participation to more central roles in collaborative practices. This view underscores that learning is relational and forward-looking, oriented toward the kinds of person learners are becoming and the roles they aspire to inhabit in society \cite{barnett2007}.

\textit{Strategies for AI companions.} To embody these principles, AI companions should be designed to situate learning in authentic contexts and scaffold the development of professional identity. They can frame problems through real-world scenarios, case studies, or discipline-specific practices, helping learners connect abstract concepts to concrete applications. Building on experiential learning cycles, AI companions can prompt learners to reflect on prior experiences, test new ideas, and refine understanding through iterative experimentation. They can also model or simulate aspects of signature pedagogies so that learners practice the distinctive ways of thinking and acting valued within their field. Finally, AI companions can support learners in navigating threshold concepts by identifying sticking points, reframing misunderstandings, and encouraging persistence through transformative moments of learning. 

While these principles guide how students learn with AI companions, effective partnership also requires that the AI itself learns about the student — constructing models, adapting behaviour, and evolving over time.

\section{The Adaptive Foundation: The AI Learns about the Student \label{sec:ai-learns-about-student}}
For learning companions to support learners effectively, they must themselves be capable of learning about the learner. This adaptive dimension concerns how AI captures, models, and interprets learner data to tailor guidance, feedback, and sequencing, and how it evolves through continual evaluation. Building on four decades of research in adaptive and intelligent tutoring systems, this section outlines a four-stage adaptive cycle, \textit{Capture, Model, Adapt,} and \textit{Evolve}, that enables AI companions to transform raw learner interactions into personalised and pedagogically grounded support.

\subsection{Capture: Recording Learners’ Digital Footprints}
\label{sec:capture}
A defining advantage of digital learning environments is their ability to record nearly every learner action, creating detailed logs of clicks, keystrokes, problem-solving steps, time spent, and communication. Decades of research in learning analytics and educational data mining have established that these fine-grained digital traces can serve as meaningful proxies for underlying learning processes \cite{papamitsiou2014learning}. The broader challenge, of course, lies in making sense of this deluge of raw data. There is often a gap between low-level interaction data and the high-level constructs they reflect. In other words, while we can capture what learners do, interpreting those actions in terms of learning gains, strategies, or misconceptions requires careful modelling and remains an ongoing focus of research \cite{pozdniakov2023using}.

This challenge is multiplied by advances in sensing and multimodal interfaces that now allow the capture of an even richer array of learner signals. \textit{Multimodal learning analytics} (MMLA) integrates data from sources such as eye-tracking, gaze, facial expression, speech, and physiological sensors (e.g., heart rate) with traditional clickstream logs. By combining these channels and the use of AI, researchers can form a more holistic picture of learning, linking observable behaviours with internal cognitive and affective states.

The addition of LLMs introduces significant new affordances in this capture phase \cite{khosravi2025generative, li2024bringing}. Unlike legacy platforms that log only discrete events, GenAI-enabled environments capture rich, open-ended learner–AI dialogues, including prompts, explanations, reflections, and feedback. These interactions form a new class of trace data—naturalistic, conversational, and embedded in process—which can reveal student epistemic frames, strategic decisions, and cognitive engagement in ways traditional logs cannot \cite{lai2025leveraging}. Furthermore, GenAI systems themselves generate metadata—rationale, summaries, uncertainty scores—that can be incorporated into LA pipelines to enrich learner models and support interpretability. 

\textit{Strategies for AI companions.} Implementing the capture phase with AI companions involves integrating traditional data collection methods with new generative capabilities. Learning management systems and educational data mining techniques reliably log structured data—quiz scores, time on task, or resource views—while LLMs help interpret unstructured learner outputs, such as free-response answers or forum posts, extracting key themes, misconceptions, or even sentiment (e.g., frustration or confidence) \cite{jin2025feedback}. Machine learning classifiers can monitor behavioural signals like idle time or erratic clicking, while AI with speech or vision capabilities can infer affective states from tone or facial expression during video-based learning. These traces should be collected intentionally and ethically in line with established privacy and analytics frameworks \cite{drachsler2016delicate}. Together, these signals form the epistemic foundation of the dual-loop framework: they allow the companion to personalise support, foster reflection, and evolve with the learner. In this sense, capture is not a passive logging activity but an active design layer that shapes how adaptivity is realised in practice.

\subsection{Model: Building a Multi-Dimensional Learner Model}
Once data are captured, the system’s next task is to model the learner, constructing a representation that encapsulates the student’s current state of knowledge, skills, strategies, and affective dispositions. Research in student modelling since the 1990s has explored how learner data can be used to drive intelligent adaptation in learning systems \cite{vanlehn2013student}. This work has yielded models capable of detecting a wide range of learner dimensions: which skills students know \cite{pelanek2017bayesian}, how long their memory of something learned will last \cite{pavlik2021logistic}, what self-regulated learning strategies they employ \cite{molenaar2023measuring}, and whether they are engaged or disengaged at any given moment \cite{baker2013assessing}. Although these models were rarely implemented at scale before the emergence of large language models (LLMs), with the notable exception of knowledge models used to decide when to advance the student \cite{koedinger2006cognitive}—they provide a strong foundation for informing the decisions of modern AI systems.

A key example is \textit{knowledge tracing} \cite{Corbett1995}, which encodes what a student currently knows relative to a domain model, a structured representation of the concepts, skills, or knowledge components that define a subject. While LLMs can be prompted or fine-tuned to diagnose specific student errors or provide feedback for incorrect solutions \cite{stamper2024enhancing}, they are less effective at inferring a student’s overall level of mastery for a skill. This is because such inference requires integrating data across all prior interactions, weighting recency and frequency, and estimating the probability of mastery \cite{ghosh2020context}. Fortunately, algorithms already exist to perform these probabilistic computations, and hybrid approaches that combine LLMs (for content interpretation and response analysis) with traditional knowledge-tracing algorithms (for longitudinal integration) have been shown to outperform either method alone \cite{scarlatos2025training}. A similar synergy applies to memory algorithms, which can predict when a student will forget previously learned material and identify optimal review intervals \cite{pavlik2021logistic}.

Decades of research on other aspects of learner modelling—such as strategies, engagement, and emotion—have also produced algorithms that can provide actionable information to LLMs. The LLM does not need to replicate these specialised functions; instead, outputs from such models can be fed into the LLM as structured text or numeric representations. In some cases, such as analysing student writing or explanations, the LLM may outperform earlier approaches, while in others—such as evaluating behavioural logs or facial expressions—dedicated machine learning models remain superior. The most effective approach combines the interpretive strength of LLMs with the precision of domain-specific models, producing a richer, more comprehensive picture of the learner.

Despite these advances, previous generations of ITSs often struggled to leverage student models for more than a single adaptive purpose within one system \cite{baker2016}. The most common application was mastery determination, deciding whether a student should move to the next topic \cite{koedinger2006cognitive}—while others focused on topic sequencing \cite{cosyn2021practical} or affective support \cite{woolf2009affect}. This limitation likely stemmed from the human design effort required to coordinate multiple adaptive dimensions. \textit{AutoTutor}, one of the most comprehensive ITSs, explored many data-driven forms of adaptivity \cite{nye2014autotutor}, yet it required decades of collaboration across dozens of researchers and typically studied one form of adaptation at a time. 
Emerging AI systems can now reason about unique combinations of learner information, experiment with alternative instructional strategies, and update their internal models based on outcomes. Just as Reinforcement Learning from Human Feedback (RLHF) has driven the development of general-purpose chatbots, \textit{Reinforcement Learning from Human Learning} (RLHL) \cite{doroudi2019s} may enable the next generation of generative AI-powered tutoring systems—ones that continually learn from interactions with students to refine their own adaptive policies.

\textit{Strategies for AI companions.}  
Building on these foundations, AI companions should adopt hybrid modelling architectures that integrate the interpretive capabilities of LLMs with established learner-modelling algorithms. LLMs can analyse open-ended learner responses to extract conceptual misunderstandings, affective cues, and evidence of metacognitive behaviour, while probabilistic or mathematical models such as knowledge tracing, memory modelling, and engagement detection integrate these signals across time. This fusion supports multi-dimensional learner models that evolve dynamically with the learner. To enhance transparency and trust, AI companions should make these models visible and interpretable through open learner model interfaces, allowing both learners and instructors to inspect, question, and adjust the AI’s inferences. Over time, such systems can employ reinforcement learning approaches to optimise how learner data are used to personalise feedback and adapt instruction, creating companions that learn not only about the student but also from them, continually refining their pedagogical strategies to maximise growth, equity, and understanding.

\subsection{Adapt: Harmonising Personalised Guidance with Shared Regulation}

Adaptive and intelligent tutoring systems are designed to modify their pedagogical decisions dynamically in response to individual learner characteristics (e.g.,  prior knowledge, affect, and engagement) and real-time learner actions \cite{aleven2016}. A classical way to conceptualise adaptivity is through two interrelated “loops” \cite{VanLehn2006}. The \textit{outer loop} operates at the macro level, managing the selection and sequencing of learning tasks. Its main responsibility is to determine which activity or problem the learner should tackle next, effectively tailoring the curriculum pathway to the individual’s progress and mastery level. The \textit{inner loop} functions at the micro level within a single task, providing step-by-step support as the student works through a problem. This includes checking correctness, offering error-specific feedback, providing hints, and scaffolding problem-solving processes \cite{aleven2016,Corbett1995}. Together, these two layers enable an ITS to adapt both \textit{what} content a learner encounters (macro-adaptation) and \textit{how} feedback and guidance are provided during task execution (micro-adaptation). For example, an outer loop might decide to reteach a struggling student a particular concept using an easier problem, while the inner loop on that problem provides immediate hints and feedback as the student attempts it.

A substantial body of evidence has demonstrated the effectiveness of adaptive learning systems relative to both classroom instruction and non-adaptive educational technologies \cite{anderson1995cognitive,vanlehn2011,Ma2014}. These systems improve cognitive efficiency and learning outcomes, yet this automation often comes at a cost. When the system assumes full control of pacing, sequencing, and support, students may become passive recipients of guidance, offloading metacognitive control and losing opportunities to develop self-regulation \cite{molenaar2022concept}. In fully automated tutors, learners have little reason to set their own goals, decide which topics to review, monitor their understanding, or plan their study, as the system performs these functions on their behalf. 

To counteract this limitation, recent research has shifted toward designs that integrate principles of socially shared regulation of learning \cite{panadero2015socially} and hybrid human–AI co-regulation \cite{molenaar2022concept}. Rather than fully automating adaptivity, these systems invite learners to share control over decision-making. Transparency mechanisms, such as Open Learner Models (OLMs), make the system’s inferences visible, prompting learners to reflect on their understanding and progress \cite{bull2020there}. Similarly, adaptive systems can embed prompts that involve learners in regulatory choices, such as asking whether they would like a hint or prefer to continue independently \cite{molenaar2019towards}. Such participatory features engage learners in monitoring and control, preserving the efficiency of adaptivity while nurturing metacognitive growth \cite{abdi2020complementing}. Goal setting plays a crucial role in this process: it anchors monitoring and self-assessment, ensuring that learners can evaluate progress against explicit objectives \cite{panadero2017review,winne2000measuring}. Without clear goals, students often struggle to activate their internal “monitoring and control” loops effectively \cite{winne2013nstudy}.  

\textit{Strategies for AI companions.}  
The rise of LLMs introduces powerful new affordances to extend and humanise adaptivity within learning companions. Traditional adaptive systems relied on a fixed pool of pre-authored problems, hints, and explanations, constraining flexibility and responsiveness. In contrast, generative AI enables dynamic co-construction of learning activities, allowing the outer loop to operate interactively in dialogue with the student. Learners’ curiosity, goals, or misconceptions can directly influence the next problems or explanations generated.  Another key affordance is explainability: generative AI tutors can articulate the rationale behind their instructional decisions in natural language. 
Whereas earlier adaptive systems could often not articulate why they evaluated a student in a certain way, LLM-based companions can express reasoning such as, “Your quick reading of high-level help resources and subsequent correct answer, combined with past correctness, suggests that your error was based on forgetting a function name rather than something deeper. 
This transparency fosters trust and self-reflection, aligning with research on learning analytics dashboards and Open Learner Models that visualise learners’ knowledge states to promote metacognitive awareness and agency \cite{bull2020there}.  
Beyond transparency, generative AI companions can actively scaffold self-regulated learning through reflective dialogue—asking, for example, “What is your goal for this session?”, “How confident do you feel about this topic?”, or “Can you explain how you solved that problem?”. Such conversational prompts have long been recognised as effective in eliciting planning, monitoring, and self-assessment behaviours. Through these exchanges, the AI companion not only adapts instruction but also cultivates the learner’s capacity to themselves adapt. This represents a significant philosophical shift: from adaptivity being something \textit{done to} the learner, where the system unilaterally controls pacing and feedback, toward adaptivity \textit{done with} the learner, in which AI acts as a co-regulator and partner in metacognitive growth, balancing personalisation with shared regulation.

\subsection{Evolve: Continuous Improvement Through Feedback Loops}
The final stage of the adaptive cycle is to evolve the system itself: evaluating how well its adaptations are working and refining them for continual improvement. Essentially, the system uses feedback not only to help the student but also to help itself, learning from its own successes and failures. In the context of adaptive learning, this concept has been formalised as \textit{design-loop adaptivity} or \textit{closed-loop evaluation}, where learner data are used to iteratively redesign and improve the instructional system \cite{aleven2016}. 

Without this evaluative phase, adaptive systems risk stagnation: their models and interventions may remain static despite shifts in curricula, contexts, or the learners themselves. The evolve phase therefore completes the adaptive cycle by treating the AI as a learning entity—one that continuously refines its instructional policies based on empirical evidence rather than fixed design assumptions. This adaptive meta-learning ensures that the system’s pedagogical foundations remain robust, equitable, and contextually responsive. In other words, the AI not only models the learner but also optimises how it supports learning.

\textit{Strategies for AI companions.} To operationalise the evolve phase, AI companions should treat instructional decisions as hypotheses to be tested rather than fixed policies to be applied. At the system level, controlled experiments and multi-armed bandit algorithms can dynamically allocate learners to instructional conditions shown to be more effective, improving both learner outcomes and system understanding over time \cite{rafferty2019statistical}. When a pedagogical strategy consistently fails to improve performance, this serves as a diagnostic signal to modify or replace it, an iterative process analogous to formative evaluation in education. Larger-scale data enable the system to go beyond learning what works to learning what works for whom and under which circumstances, supporting equitable as well as effective adaptation. Simulated learners that model realistic student behaviours offer a complementary approach, allowing new instructional strategies to be tested rapidly before deployment \cite{gao2025agent4edu}. A more recent innovation leverages generative AI as a co-designer: LLMs can propose new feedback prompts, scaffolding sequences, or dialogue structures that are then validated through empirical testing with real learners. Together, these strategies ensure that the AI companion does not merely adapt to the learner in the moment but continuously improves how it supports learning over time.

\section{The Responsible Design Foundation: AI Acts with Integrity and Upholds Human Values} \label{sec:ethical-companions}

The highly personalised and embedded nature of learning companions means that their influence extends well beyond technical optimisation. As companions capture rich learner data, shape study habits, and mediate everyday decisions about effort, strategy, and assessment, they also carry social, ethical, and institutional consequences. Responsible design is therefore not a peripheral concern but a core requirement: learning companions must protect learners from harm, make their operations intelligible, remain under meaningful human oversight, and promote equity rather than reproduce existing disparities. 

Trustworthiness is the unifying goal of this foundation, encompassing the expectations of students, educators, technologists, institutional leaders, and where relevant, parents and caregivers. We organise this around four interwoven commitments: \emph{security}, \emph{transparency}, \emph{accountability}, and \emph{inclusion}. While these commitments apply to any AI application, we clarify how they play out distinctively with learning companions, and the specific choices that platform owners must make about the data they capture, how it is used, and by whom.

\subsection{Security: Protects privacy, data, and digital safety through lawful and responsible AI use}
UNESCO guidance emphasises that AI systems must be mindful of data ownership, data privacy, and data availability for the public good, and should adopt principles of privacy and security by design \cite{miao2021ai}. Without robust safeguards, the rich interaction data generated by learning companions could be misused for unintended profiling, surveillance, or exposure of private information. Secure practice includes separating personally identifiable information from interaction data, preferring local or institutionally hosted processing for sensitive tasks, and implementing tiered retention with auditable deletion logs.

The interaction logs and transcripts generated by AI companions can range from relatively innocuous content, such as problem-solving traces in an algebra tutor, to sensitive personal disclosures, such as emotional reflections on a difficult work placement shared with a conversational companion. Whether educational institutions are licensing commercial products hosted remotely, deploying on-premises, or adopting a hybrid approach, they must make deliberate choices about how much data they archive and for what purposes.

Security is therefore not only technical but extends to psychological safety. Given the social and conversational nature of AI companions, learners may disclose more than they intend or become emotionally dependent on interactions in ways that require careful design. Companions should protect learners' digital wellbeing by avoiding manipulative designs, implementing age-appropriate guardrails, and preventing harmful or crisis-related content from bypassing safety layers. Robust defences against prompt injection, content exfiltration, and adversarial misuse are essential, as are proactive risk-scanning mechanisms that detect anomalous system behaviour. By treating data protection and wellbeing as preconditions for adaptivity rather than trade-offs, security becomes an enabler of trustworthy personalisation.

\textbf{Strategies for AI companions.} To operationalise security and safety, learning companions should employ modular architectures that isolate sensitive processes, adopt privacy-preserving representations such as derived or aggregated features rather than raw traces, and automate privacy notifications so learners understand what is stored and why. Synthetic or differentially private data can be used for model updates where possible. Periodic self-checks, anomaly detection pipelines, and red-team testing ensure secure operation across the learning lifecycle. Companions should align their prompts, retrieval, and sensitive content filtering with institutional academic-integrity settings, and maintain stable, versioned model deployments so that changes do not invalidate the integrity of established companion behaviour.

\subsection{Transparent: Makes AI processes visible, explainable, and open to understanding}
The literature consistently underscores that transparency and explainability are not optional add-ons but foundational requirements for AI-powered educational tools. These requirements stem from two interrelated imperatives: establishing calibrated trust and enabling meaningful, pedagogically actionable insight for all stakeholders—students, educators, and institutional decision-makers. When AI systems make decisions that influence learning pathways, assessment outcomes, or resource recommendations, opaque processes undermine confidence. Learners may over-trust fluent but unjustified suggestions, while educators may resist adoption altogether when they cannot see why or how the system arrived at its outputs. This tension is especially acute for complex “black-box’’ architectures, such as deep neural networks, whose internal logic is unintelligible without deliberate design for interpretability.

Explainable AI (XAI) frameworks address this challenge by making the system’s reasoning visible and comprehensible. In education, however, explanation must be pedagogically purposeful, not merely technically informative. The XAI-ED framework \cite{khosravi2022} formalises this insight, arguing that explanations should actively advance teaching and learning—for instance, by helping students understand misconceptions or by enabling educators to diagnose why a learner model reached a particular conclusion. XAI thus supports fairness, accountability, and ethical integrity in high-stakes uses of AI—such as automated grading, risk prediction, and personalised feedback—mitigating concerns about bias and undue influence \cite{holmes2022ethics}. Ultimately, transparency and explainability ensure that AI systems operate as accountable partners within the learning ecosystem. 

In the case of AI tutors, transparency and explainability can also build directly on a long-standing tradition of Open Learner Models (OLMs) \cite{bull2020}, which make the system’s representation of a learner’s knowledge, skills, or progress available for inspection and negotiation. Classic OLM work demonstrates that simply exposing the model fosters trust, supports metacognition, and encourages learners to take a more active role in monitoring and directing their progress. Integrating XAI enables OLMs to reveal the evidential basis for inferences (e.g., “This misconception was detected because your last three attempts showed…”) and the rationale behind pedagogical decisions (e.g., “This resource is recommended because it addresses the specific pattern of errors you exhibited in…”). This transformation reframes the OLM from a passive display into an active pedagogical instrument. With transparent, explanation-rich learner models, students can interrogate AI recommendations, challenge inaccuracies, and understand the logic behind their personalised pathways. 

\textbf{Strategies for AI companions.}  
To build calibrated trust and enable meaningful learner agency, learning companions must adopt a suite of strategies that foreground transparency in both their operations and their pedagogical interactions. First, companions should clearly communicate what data they collect, how it is used, and which aspects of the learner model inform specific recommendations or feedback. This includes providing layered explanations that range from simple justifications (e.g., “I recommended this resource because…” ) to more detailed technical rationales for educators or advanced users. Second, companions should make their reasoning processes visible by revealing the evidence, inference steps, or patterns that shaped their conclusions, allowing learners and teachers to interrogate, contest, or refine the model’s understanding. Third, companions should openly disclose uncertainties and limitations—highlighting when an inference may be incomplete or when data is insufficient—to prevent overconfidence and encourage critical engagement. Fourth, transparent companions should make their capabilities, constraints, and guardrails explicit, helping learners calibrate expectations and avoid inappropriate reliance on AI-generated suggestions. Finally, transparency must extend to the evolution of the learner model over time, showing how new actions or responses update the system’s interpretation of progress.

\subsection{Accountable: Retains human oversight, reliability, and the right to question AI-supported decisions}

Accountability ensures that companions remain corrigible, governable, and aligned with educational responsibility. Even when models are highly capable, educators and institutions, not algorithms, retain responsibility for learning experiences and outcomes. This requires meaningful human oversight, reliable performance, and clear avenues for contesting AI-supported decisions \cite{whittlestone2019, floridi2023}.

High-stakes inferences, including mastery promotion, at-risk flags, and achievement predictions, should incorporate human-in-the-loop checkpoints. Learners and teachers require channels to dispute model states or recommendations, and these disputes should feed into refinement pipelines or policy-rule updates. Reliability depends on continuous monitoring of hallucination rates, harmful suggestions, and model drift, complemented by retrieval-augmented generation, domain-validated templates, and specialised models such as knowledge tracing that constrain LLM free-form generation. Audit logs linking inputs, model states, actions, and outcomes enable rigorous incident analysis.

Crucially, accountability must be learning-aligned. Companions should be evaluated not only on output accuracy but on their impact on retention, transfer, metacognition, and durable learning \cite{bjork2013, soderstrom2015}. This connects directly to the paper's central argument: a companion that produces high task performance while undermining genuine learning has failed its core purpose, regardless of how technically reliable it appears. Ensuring that adaptivity enhances rather than bypasses productive struggle is therefore a core accountability criterion, not a secondary consideration.

\textbf{Strategies for AI companions.} Companions can include review modes for educator verification of system actions, implement dispute workflows for learners, and maintain auditable decision trails. They should run continuous reliability benchmarks and use fallback behaviours, such as simplified rule-based responses, when confidence drops. Governance reviews and incident post-mortems ensure that companions remain reliable, correctable, and aligned with educational goals. Evaluation frameworks should incorporate delayed retention and transfer measures rather than relying solely on immediate performance metrics.

\subsection{Inclusive: Designs AI that is fair, accessible, and responsive to diverse learners}

Inclusion requires that learning companions expand opportunity rather than personalising disadvantage. Because models are trained on data reflecting historical inequities, they risk reproducing bias in content, tone, and adaptive decisions. Inclusive companions therefore embed algorithmic fairness, accessibility, and cultural responsiveness at every stage of design and deployment.

Datasets and outputs should be audited for representational harm and performance disparities across demographic groups \cite{birhane2021,buolamwini2018,rajkomar2018}. Fairness-aware training and post-processing can mitigate some risks, but inclusive design must also attend to interactional accessibility and contextual appropriateness. Following Universal Design for Learning principles, companions should support multiple means of engagement, representation, and expression—adjustable reading levels, multimodal explanations, captioning, multilingual support, alternative input modes \cite{meyer2014}. 

Supporting low-bandwidth and mobile-first environments is essential to prevent digital exclusion. Participatory co-design with under-represented learners and educators ensures that prompts, examples, scaffolds, and help policies reflect local curricula and cultural norms \cite{bulathwela2024}. Finally, inclusive companions must counter risks of cognitive offloading by providing staged hints, self-explanation prompts, and faded support so that help develops learner autonomy rather than dependency \cite{risko2016}.

\textbf{Strategies for AI companions.}  
Inclusive companions can dynamically adjust language complexity, offer multilingual and multimodal options, and apply fairness-aware personalisation that tracks disparate effects. Accessibility-first modes can prioritise simplified layouts, alternative inputs, and sensory-friendly feedback. Co-design cycles with diverse learners help ensure grounding in real contexts. Fairness dashboards, usability audits, and localisation reviews maintain inclusivity as systems evolve.

\section{Case Studies of LLM-Powered Tools Designed to Support Learning}\label{sec:case-studies}

This section presents five case studies of LLM-powered educational systems intentionally designed to support student learning across diverse contexts and disciplines. Each case study is authored by researchers who led or made significant contributions to the development of the tool described, providing firsthand insight into the design decisions, empirical findings, and practical challenges involved. 

Section~\ref{sec:khanmigo} presents Khanmigo, an AI-powered tutor embedded within the Khan Academy platform, designed to scaffold skill practice through guided hints, worked examples, and cognitive engagement measurement across a broad global curriculum. Section~\ref{sec:ripple} presents RiPPLE, a university-scale learning platform that integrates generative AI companions into cycles of student content creation, peer evaluation, and personalised practice, supporting active and social learning at scale. Section~\ref{sec:codehelp} presents CodeHelp, which provides scalable, solution-free programming support through structured queries and guided tutoring workflows, modelling productive struggle and effective help-seeking as a metacognitive skill. Section~\ref{sec:jeepyta} presents JeepyTA, an AI teaching assistant embedded in course discussion and feedback workflows, supporting dialogue, formative feedback, and brainstorming while preserving configurable instructor oversight. Finally, Section~\ref{sec:recast} presents Recast, an institutional platform developed at the University of Technology Sydney that enables educators to design and deploy steerable, context-sensitive AI assistants grounded in course-specific resources, with full enterprise governance and deliberate attention to responsible design at scale.
For each tool, we describe its purpose and deployment setting and analyse how its design aligns with the pedagogical, adaptive, and responsible foundations articulated earlier in the paper. Collectively, these cases illustrate how generative AI can move beyond task-oriented assistance toward learner-centred support that promotes understanding, reflection, and sustained engagement across a wide range of educational contexts.


\subsection{Khanmigo: An AI-Powered Tutor Integrated into a Global Learning Platform} \label{sec:khanmigo}
\subsubsection{Overview of Khanmigo}
Khan Academy is a supplemental online instruction and practice platform meant to provide students with the independent practice opportunities they often do not receive with their core curriculum. Khanmigo is an AI-powered tutor integrated into this practice platform to provide scaffolding and feedback in ways not available prior to the advent of generative AI. In the 2025-26 school year, approximately 1M students and teachers will engage with Khanmigo in classrooms as part of the Khan Academy Districts offering. Research across multiple datasets from multiple digital learning platforms has demonstrated that each practice opportunity for students results in predictable gains toward skill mastery \cite{koedinger2023astonishing}. Unfortunately, students do not consistently get the opportunity to engage in the amount of practice needed to reach mastery. Khan Academy and Khanmigo are meant to change that.

\begin{figure}[t]
    \centering
    \begin{subfigure}[t]{0.65\textwidth}
        \centering
        \includegraphics[width=\textwidth]{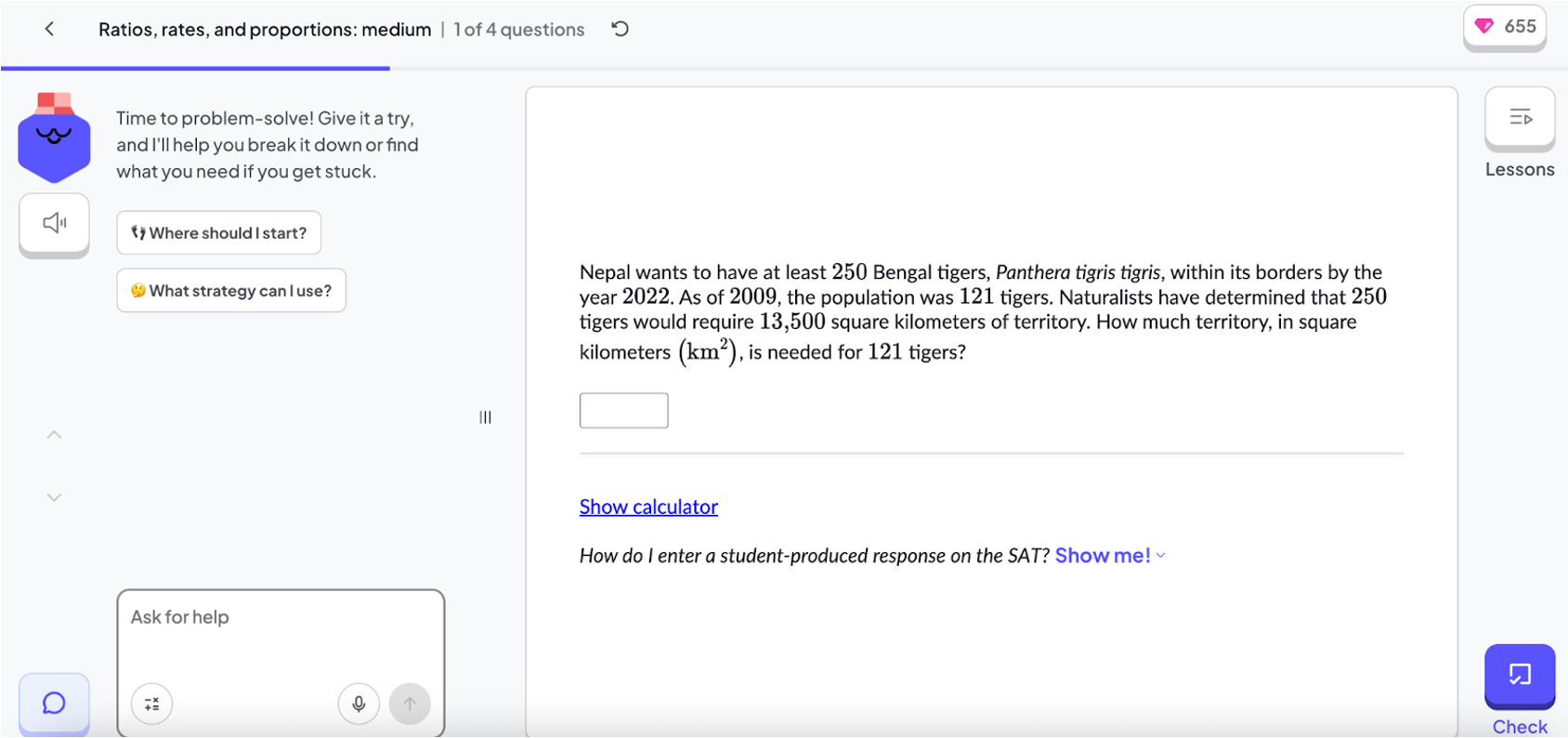}
        \caption{Khanmigo in companion mode working with a Khan Academy question.}
        \label{fig:khanmigo1}
    \end{subfigure}
    \hfill
    \begin{subfigure}[t]{0.30\textwidth}
        \centering
        \includegraphics[width=\textwidth]{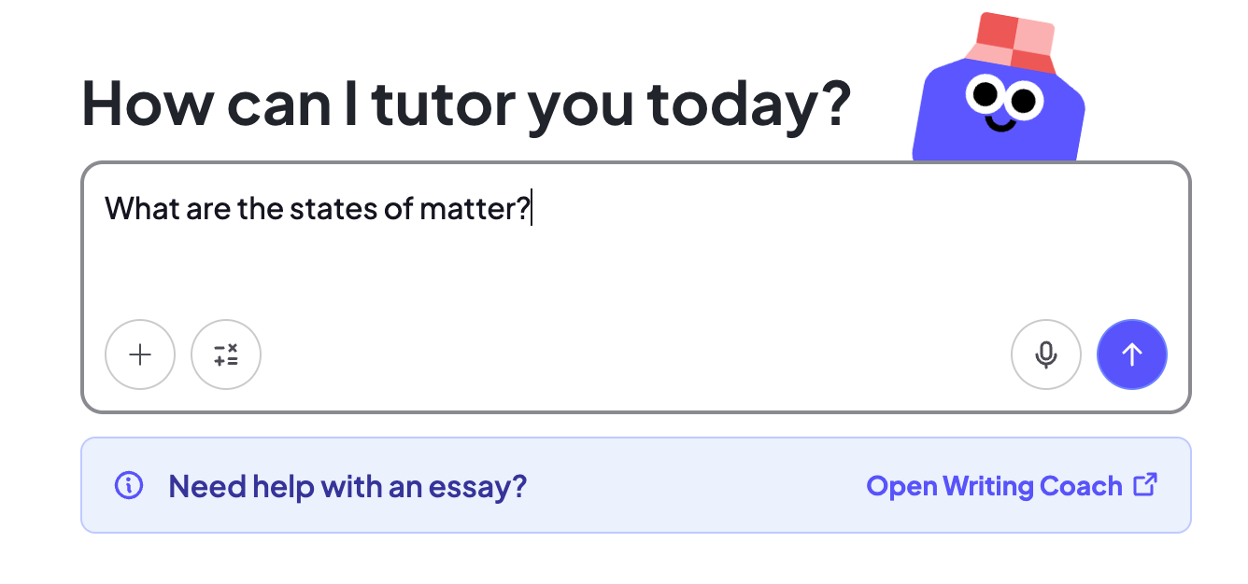}
        \caption{Khanmigo in TutorMe mode working on a question the student inputs.}
        \label{fig:khanmigo2}
    \end{subfigure}
    \caption{Examples of Khanmigo operating in different interaction modes. Examples of Khanmigo operating in different interaction modes. The left image shows companion mode, in which Khanmigo supports a student working through a Khan Academy problem with hints and guided reflection. The right image shows TutorMe mode, in which students bring their own questions and problems to the system}
    \label{fig:khanmigo_modes}
\end{figure}

Learners interact with Khanmigo in a number of ways. First, Khanmigo is available for students as they are working on Khan Academy content (see Figure~\ref{fig:khanmigo1}), in what Khan Academy calls companion mode (not to be confused with the term companion used elsewhere in this paper). Second, students can bring outside questions and problems to Khanmigo in TutorMe mode (see Figure~\ref{fig:khanmigo2}). Third, students can interact with Khanmigo in the essay coach feature, where Khanmigo provides feedback and coaching to students on their writing assignments. Finally, there are 20+ other distinct activities where learners interact with Khanmigo in specific ways, for example in a debate, to play a vocabulary game, or in co-creating a story. For the purposes of simplicity, we will not address the latter category in this paper. 

\subsubsection{Pedagogical Foundations of Khanmigo}
Khanmigo is not designed to be a complete learning companion. As such, it does not meet all of the pedagogical foundations described above and is not intended to. Rather, it is designed explicitly to do what is needed to enhance the practice in Khan Academy’s system, mainly by providing improved scaffolding and feedback during practice.

\textbf{Focus on student cognitive engagement}.
The initial efforts to design Khanmigo focused on the tutor moves it would make, drawing on literature about what good human tutors do (e.g., \cite{graesser1995collaborative}). It became clear that Khanmigo could in fact replicate good tutor moves. However, the real problem to tackle was student cognitive engagement. Examination of transcripts revealed too many students simply typing “I don’t know,” “idk,” and other generally passive responses. In order to address this issue, there was a need for a way to measure the cognitive engagement of students. Chi and Wylie’s \citeyear{chi2014icap} ICAP framework provided a helpful taxonomy by which students’ chats with Khanmigo could be categorized as passive, active, or constructive \cite{weatherholtz2025cognitive}. An LLM judge was created to do this categorization automatically. Initially, the findings of the categorization were disappointing. Students were passively engaged at levels that meant Khanmigo was not likely to improve learning outcomes for them. However, having the measure meant the Khan Academy team could test the impact of changes, like those described below, on engagement, and are now able to adjust Khanmigo behavior to optimize for student cognitive engagement. 

\textbf{The walkback of socratic tutoring}. 
The very first prompt ever made to a pre-release version of GPT-4 by Khan Academy staff started with the direction, “You are a socratic tutor.” Much early effort in prompt engineering went into preventing the tutor from giving answers, and creating evaluations to monitor whether it did. However, review of transcripts of students interacting with Khanmigo and a re-read of research literature led to a change in approach. Transcripts revealed significant frustration in students who legitimately did not know content when Khanmigo continued to just ask question after question. This led to the abandonment of interaction with Khanmigo rather than learning. Reviewing the intelligent tutoring literature revealed studies that seemed to suggest that when faced with a problem, if a student has any idea how to solve it, they should make an attempt. If they do not, they should get a hint \cite{aleven2004toward}. Note that a hint is different from a socratic question. In response, the team redesigned Khanmigo in companion mode so that it encourages students to make an attempt and provides hints when needed. Then, if the learner’s attempt is incorrect, Khanmigo engages deeply with the student on the problem, encouraging reflection on where they went wrong, including providing the answer in the form of a worked example. 

\subsubsection{Adaptive Foundations of Khanmigo}
The capture and modeling of student proficiency has long been a feature of Khan Academy (prior to the introduction of Khanmigo). The modeling of student proficiency is done at a much more simplistic level than modeling in many intelligent tutors. Students meet clear heuristic rules to move their status on each skill from attempted to familiar to proficient to mastered. For example, when working on an exercise, getting 100\% results in proficient status. Getting questions on that skill correct on a mixed practice assignment, like a unit test or course challenge, bumps that status to mastered. Students are able to see their current status on every skill in their course (or other courses they wish to explore). At one time, Khan Academy had a more probabilistic algorithm working behind the scenes to both estimate mastery and make recommendations. Students vociferously shared that they did not understand what they needed to do to get to mastery; they wanted to have a clear understanding of how many questions on what assignments they needed to get right to move up. This is understandable if their homework assignment is “get to mastery on systems of equations” rather than “complete all the questions in the exercise” which is typical in many curricula. In the same vein, teachers did not understand why the system was making particular recommendations, and many of the recommendations did not align with the curriculum being taught in class. So, the system operates on clear rules and students are assigned groups of skills in units on which to get mastery, often by their teacher rather than the system.

The mastery system in Khan Academy provides a basis for adaptivity in conversations with Khanmigo. When a student is working on a problem with Khanmigo in companion mode, the system inserts the students’ mastery level on that skill and pre-requisite skills into the prompt. This has shown promising results in improving student success on future problems on that skill. Khan Academy also experimented with collecting student interests and injecting them into prompts to influence conversations but has not resulted in improvements to student engagement or learning to date. For all the interest in personalization, so far Khanmigo has been most successful with information about how much the student knows about the current skill being practiced. 

\subsubsection{ Responsible Design Foundations of Khanmigo}
Given Khanmigo is largely deployed with learners in U.S. grades 3-12 in schools, significant attention was paid to safety and security. 

\textbf{Moderation and transparency of student conversations}. Khanmigo involves minors interacting with LLMs, so there was a need for visibility into their conversations. For students in classrooms, their entire conversations are viewable by their teacher (and conversations in individual accounts are visible to parents). Student responses are all sent through a moderation AI that flags instances of violence, hate, self-harm, sexual content, and other concerning issues. If a response meets the criterion, it is flagged to the teacher or parent and Khanmigo redirects the conversation. There is a trade-off in the system between catching all concerning interactions and false positives. Some conversations, like discussions of the deaths in Romeo \& Juliet, are generally going to cause flags. The balance is to make sure that teachers are not receiving so many flags that they begin to ignore them. 

\textbf{Data privacy and security}. Khan Academy maintains agreements with the foundational model providers it uses stating that no user data will be used to train their models. The agreements also contain clear data deletion policies where data are only held by the model providers long enough to ensure there are no concerns about safety, and then they are deleted. Similarly, Khan Academy has data retention policies that outline when chats will be deleted, given data deletion is the best guarantee against breaches.

\subsection{RiPPLE: An AI-Powered Student Co-Creation and Peer Learning Platform} \label{sec:ripple}

\subsubsection{Overview of RiPPLE}
\begin{figure}[t]
    \centering
    \includegraphics[width=1\textwidth]{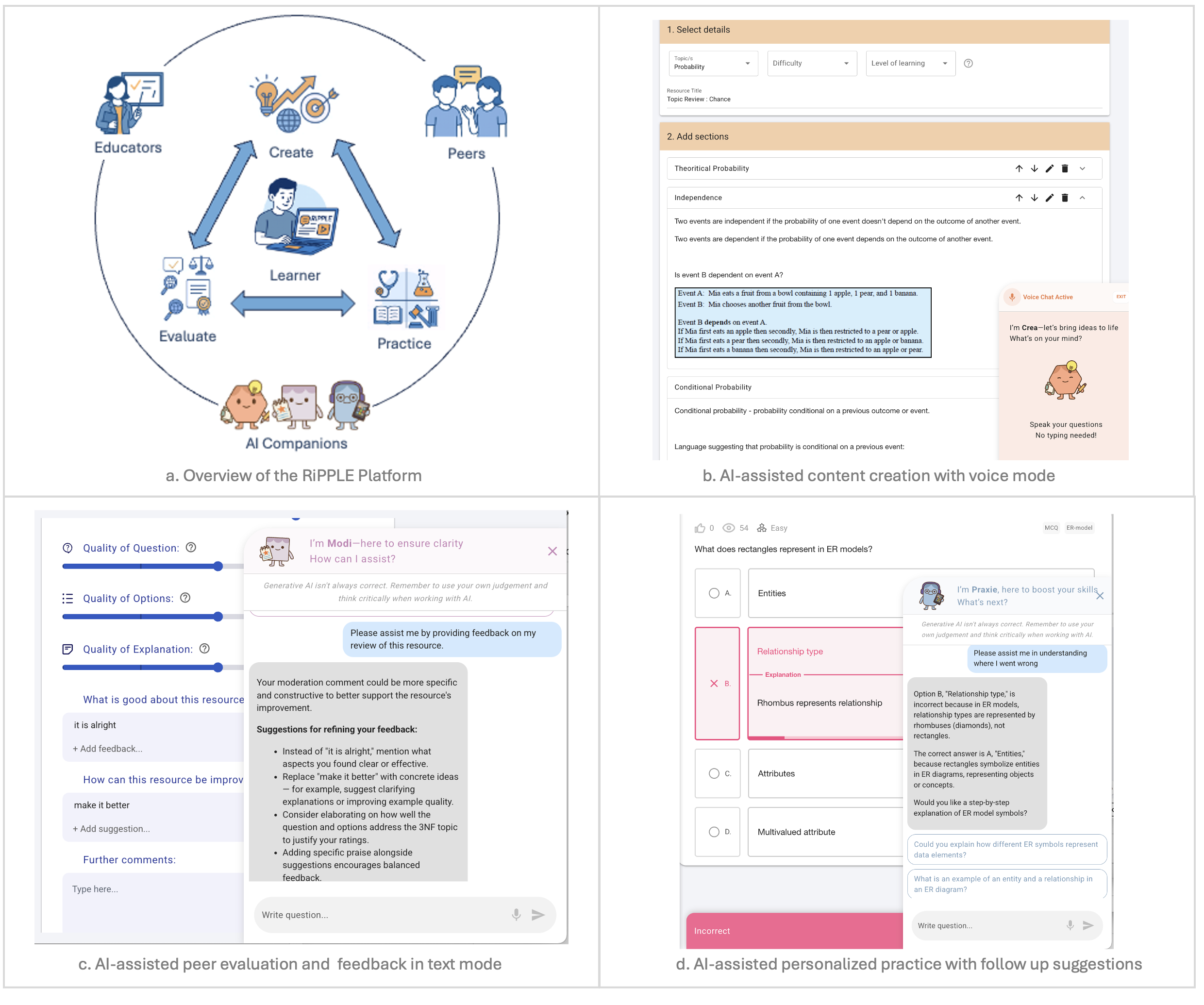}
    \caption{Illustrative overview of the RiPPLE platform and its AI-assisted learning workflows. Panel (a) presents the core learning cycle—creation, evaluation, and personalised practice—supported by educators, peers, and AI companions. Panel (b) shows AI-assisted content creation with optional voice-based interaction. Panel (c) demonstrates AI-supported peer evaluation, where learners receive constructive guidance to refine their moderation comments. Panel (d) depicts AI-enhanced personalised practice, including tailored explanations and follow-up prompts that scaffold deeper understanding. Together, these features highlight RiPPLE’s integration of generative AI to strengthen learning, feedback, and learner agency at scale.}
    \label{fig:ripple-study}
\end{figure}

RiPPLE is an AI-enhanced learning platform designed to support active, social, and personalised learning at scale. Used by more than 80,000 students across multiple disciplines, RiPPLE situates learners at the centre of a triadic cycle of creation, evaluation, and practice, supported by peers, educators, and, more recently, AI learning companions \cite{khosravi2023}. Figure~\ref{fig:ripple-study} illustrates this ecosystem and the platform’s evolving integration of generative AI.

Panel (a) depicts RiPPLE’s core workflow. Students generate bite-sized learning resources—multiple-choice questions, explanations, worked examples, or conceptual problems—functioning as ``experts-in-training''. Because student-generated resources may include inaccuracies or incomplete pedagogical reasoning, each resource is evaluated by multiple peers and, where necessary, by instructors. High-quality resources are published to a shared repository; those requiring improvement are returned to authors for revision. Across these activities, RiPPLE continuously infers learners’ topic-level abilities and recommends personalised practice resources aligned with their evolving mastery.

Building on this foundation, RiPPLE now integrates a suite of generative AI learning companions that support students across all three learning activities. Panels (b), (c), and (d) illustrate how companions provide real-time guidance during content creation, scaffold constructive peer evaluation, and enhance personalised practice through adaptive explanations, just-in-time prompts, and targeted follow-up questions. These capabilities extend RiPPLE’s capacity to provide scalable, context-aware support while preserving learner agency and pedagogical integrity.

\subsubsection{Pedagogical Foundations of RiPPLE}
RiPPLE’s design is grounded in well-established principles from the learning sciences that emphasise active knowledge construction, generative engagement, and iterative cycles of practice. Rather than positioning learners as passive recipients of information, the platform structures learning around cognitively demanding tasks that foster elaboration, retrieval, metacognition, and peer-supported reasoning. The integration of generative AI learning companions is intentionally aligned with these mechanisms: companions do not simply provide answers, but serve as prompts, scaffolds, and reflective partners that help learners think more deeply, regulate their learning, and engage productively with challenging material. The following sections outline how RiPPLE operationalises these pedagogical foundations across creation, peer evaluation, and personalised practice.

\textbf{Elaboration through content creation and peer feedback.}
A defining pedagogical feature of RiPPLE is its requirement for students to generate learning resources, a process shown to enhance student engagement, promote deeper processing, and support meaningful learning \cite{khosravi2021charting}. The cognitive effort required to explain a concept, select core ideas, anticipate misconceptions, and craft plausible distractors promotes elaboration, a central mechanism for deep learning \cite{hardy2014student}. Structured peer feedback amplifies these benefits. When reviewing peers’ contributions, learners must assess correctness, clarity, and pedagogical value, requiring them to articulate reasoning, detect conceptual gaps, and consider alternative perspectives \cite{topping2017peer}. This reciprocal process of creating and evaluating resources positions learners simultaneously as teachers and critics, strengthening their elaboration and metacognitive awareness. AI learning companions enhance both creation and peer evaluation by prompting deeper cognitive engagement. During creation, companions encourage learners to justify design choices, link concepts to prior knowledge, and refine explanations (e.g., “How does this connect to the foundational idea introduced earlier?”) \cite{pozdniakov2025ai}. During peer evaluation, companions scaffold high-quality feedback by modelling constructive comments, identifying missing rationale, and encouraging reviewers to make comments more actionable. These supports deepen cognitive engagement without replacing the epistemic work required for learning.

\textbf{Spaced retrieval and practice through personalised cycles of engagement.}
RiPPLE embeds the principles of spaced retrieval and spaced practice by recommending activities based on learners’ mastery trajectories. Retrieval practice—particularly when effortful—is one of the most powerful mechanisms for strengthening long-term memory \cite{roediger2011critical}. RiPPLE ensures that opportunities for retrieval are distributed across time by interleaving practice with cycles of content creation and peer review, enabling repeated engagement with concepts from different angles. AI companions further reinforce these mechanisms by providing adaptive, personalised support during practice. They may prompt learners to recall related principles before offering hints, encourage justification of chosen answers, or surface past misconceptions to stimulate reflection. Companions can also generate follow-up challenges, recommend revisiting earlier created or reviewed resources, or propose consolidation tasks when recurrent patterns of error are detected. These behaviours ensure that retrieval remains active, distributed, and conceptually relevant, supporting deeper consolidation and transfer.

\subsubsection{Adaptive Foundations of RiPPLE}
Central to RiPPLE’s design is the integration of learner models that drive personalised recommendations and shape AI companions’ interactions. The platform captures rich digital traces—including performance, contribution history, peer evaluations, time-on-task, and patterns of misconception—to form multi-dimensional representations of each learner’s knowledge, progress. These learner models power RiPPLE’s adaptive recommendation engine, which selects suitable practice items, identifies gaps in conceptual understanding, and schedules review activities at optimal intervals \cite{khosravi2023}. When integrated with generative AI companions, these models enable the system to deliver personalised, contextualised support. Beyond cognitive modelling, RiPPLE captures contextual features of the learning experience such as the difficulty of contributed questions, peer ratings, or the diversity of resources encountered. Learning companions use these contextual signals to determine the granularity of explanations, when to prompt reflection, or when to escalate support. This harmonises micro-level adaptivity with macro-level guidance, enabling companions to learn from and evolve with each student’s trajectory.

\subsubsection{Responsible Design Foundations in RiPPLE}
RiPPLE’s implementation of learning companions is grounded in key elements of the responsible design principles outlined earlier, with particular emphasis on transparency, explainability, and meaningful human oversight. These foundations ensure that AI support enhances learning without undermining learner agency, instructional control, or pedagogical integrity.

\textbf{Explainable learner modelling.} Aligned with the principles of XAI-ED \cite{khosravi2022}, RiPPLE enables learners to understand why the system makes particular recommendations or highlights specific areas for review. Learning companions provide clear, context-sensitive explanations of their decisions, such as: “I suggested revisiting this topic because your last three attempts showed the same misconception.” These explanations make the system’s reasoning visible and intelligible, helping learners calibrate their trust in the platform and encouraging active engagement in metacognitive monitoring. By situating feedback within the learner’s behavioural and conceptual history, explainability becomes a pedagogically meaningful scaffold rather than a purely technical disclosure.

\textbf{Instructor oversight and auditability}. Human oversight is preserved through comprehensive instructor dashboards that surface all learner-generated content, AI-supported feedback exchanges, and moderation outcomes. Educators can review resources that have been flagged for potential issues, and identify recurring misconceptions across the cohort. This auditability allows instructors to intervene when needed, refine instructional strategies, and maintain authority over judgements that carry pedagogical or assessment stakes. Importantly, the design ensures that automated systems do not make unilateral high-stakes decisions; instead, AI acts as a support layer that augments, rather than replaces, human expertise.

\subsection{CodeHelp: An AI-Powered Tutor for Scaffolded Programming Support} \label{sec:codehelp}

\subsubsection{Overview of CodeHelp}

CodeHelp is an LLM-powered web application designed to provide scalable, on-demand support to programming students without revealing code solutions~\cite{liffiton2023codehelp}.  This design is supported by recent research showing that students in programming courses value fast access to help but prefer scaffolded guidance over direct answers~\cite{denny2024desirable}.  CodeHelp is one of the flagship applications built on the Gen-Ed (Generative AI for Education) framework, an open-source infrastructure for constructing educational web applications. Gen-Ed provides authentication (including single sign-on), LMS integration via LTI, class enrollment and role management, administrative dashboards, and data export capabilities. 

\begin{figure}[t]
    \centering
    \includegraphics[width=1\textwidth]{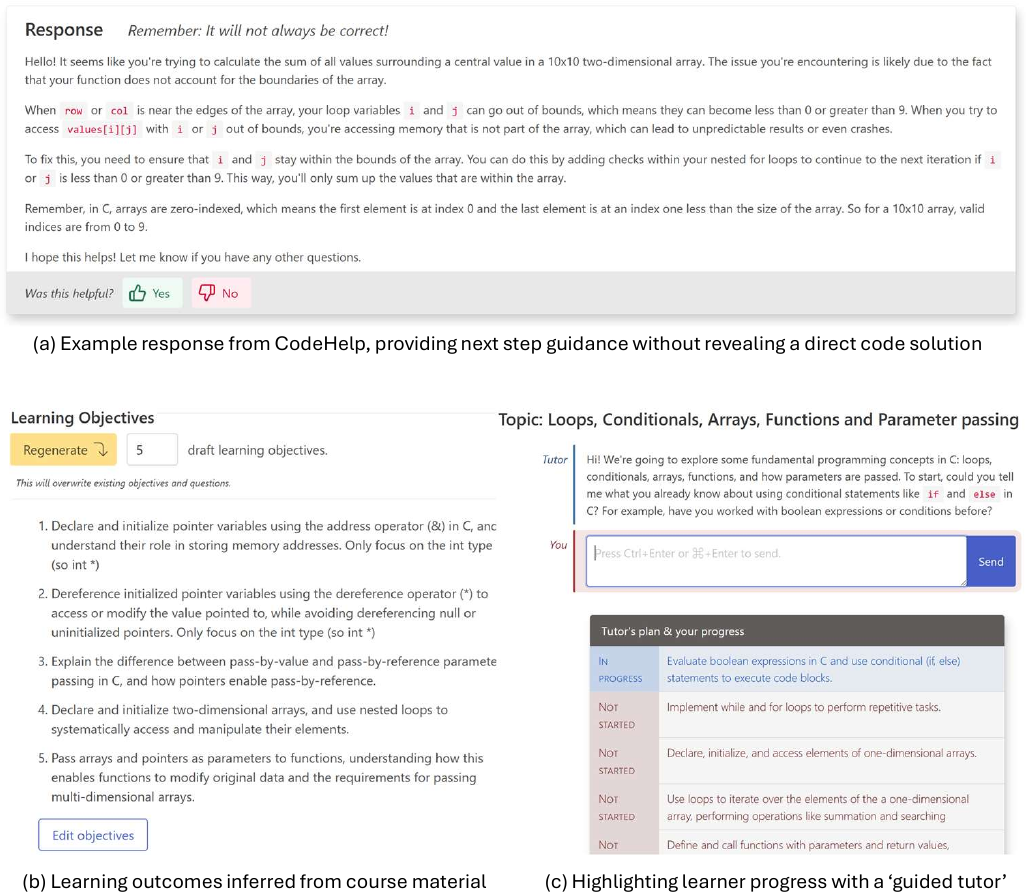}
    \caption{Illustrative examples from the CodeHelp platform. Panel (a) illustrates a guardrailed response in the `Code query' mode after a student submitted code containing a bug.  Panel (b) illustrates the instructor configuring a `guided tutor' session.  After uploading relevant course material, a set of learning outcomes are generated for review.  These can be edited directly or regenerated.  Each learning outcome is accompanied by a step-by-step associated plan (not shown in the figure), which is also generated by CodeHelp for the instructor's review, and which scaffolds the learner's interaction with the tutor.  Panel (c) shows the start of a `guided tutor' session, with a table that highlights the learner's progress through the plan.} 
    \label{fig:CodeHelpExample2}
\end{figure}

CodeHelp provides two primary forms of support. The first is the `Code Query' feature, which allows students to request help with programming issues such as debugging, interpreting error messages, or understanding conceptual aspects of their code. Students submit structured inputs that include the programming language, relevant code, error messages, and a description of their question. Responses are generated through a multi-stage prompting pipeline designed to avoid providing solution code. This pipeline includes (1) a sufficiency check to determine whether the student's query contains adequate information, (2) generation and scoring of multiple candidate responses, and (3) a post-processing step that removes any code blocks from the response. This design ensures that students receive conceptual guidance without direct answers.  An example of a guardrailed response from CodeHelp is illustrated in Figure \ref{fig:CodeHelpExample2}(a).

The second feature is the `Tutors' system. CodeHelp supports both open-ended `Inquiry Chats', where students initiate discussions about topics of interest, and `Focused Tutors', which are configured by instructors around specific learning objectives. For `Focused Tutors', instructors upload course materials and CodeHelp automatically proposes learning objectives and associated assessment questions. Instructors edit and refine these suggestions prior to deployment. During an interactive session with a student, the tutor guides the student through the objectives interactively, probing for understanding and requiring evidence of mastery before progressing.  Figures \ref{fig:CodeHelpExample2}(b) and (c) illustrate examples of the learning objectives generated from provided course material, and the beginning of a `guided tutor' session, respectively. 
  Responses are deliberately framed in a positive and encouraging tone, modelling supportive human tutoring practices \cite{lin2022exploring}. CodeHelp shares similarities with other programming assistants such as CodeAid~\cite{kazemitabaar2024codeaid}, particularly in its use of guardrails to prevent direct code generation. 

\subsubsection{Pedagogical Foundations of CodeHelp}

\textbf{Deep and Interactive Learning}
CodeHelp is intended to function similarly to a human tutor by engaging in guided, back-and-forth interactions with students rather than simply delivering answers. This approach aligns with the `Deep and Interactive Learning' principle by promoting active cognitive engagement rather than passive consumption of solutions.  With the `Focused Tutors' feature, students engage in dialogue as they step through questions posed by the tool which relate to the learning outcomes approved by the instructor.  As they do so, the system probes for understanding, asks follow-up questions, and adapts its explanations to the student's responses. This feature requires students to demonstrate understanding before they progress.  In the `Code Query' feature, students are free to ask their own questions and they receive conceptual explanations and guidance rather than complete implementations.  This requires students to interpret the guidance provided, rather than passively consume solutions. 

\textbf{Guided Scaffolding and Productive Struggle}
A defining feature of CodeHelp is its explicit avoidance of solution code when students are seeking help in the the `Code Query' mode. The multi-stage prompting pipeline that CodeHelp uses serves as a pedagogical guardrail to support understanding rather than bypassing it. The use of such guardrails also introduces an element of productive challenge, where students have to interpret the feedback provided and apply it within their own programming context.  Responses are deliberately phrased in a positive and encouraging tone, in order to sustain motivation.


\textbf{Learning to Learn and Metacognitive Development}
CodeHelp provides a query interface that is divided into four areas: programming language, relevant code fragment, error message, and free form question.  This structure is designed to ``provide guidance to students about what information is typically needed for an effective query'' and to give ``students an opportunity to practice asking technical questions, providing the necessary relevant context'' \cite{liffiton2023codehelp}.  More importantly, CodeHelp includes a `sufficiency check' as part of the prompting pipeline, which determines if all essential information for answering the query has been provided by the student.  If any important information is missing, rather than providing a potentially inaccurate repsonse, CodeHelp will respond to the student's question by asking for the missing information.  This combination of the structured input field and the sufficiency check helps students form precise questions and thus models an important aspect of effective help seeking, which is an important metacognitive skill.  

The `Focused Tutors' mode further supports metacognitive development by prompting students to reflect on their understanding as they work through instructor-defined learning outcomes. Students are asked open-ended questions that probe conceptual understanding, and they are expected to articulate their reasoning before progressing through to the next step in the guided tutor plan. When misunderstandings arise, the tutor provides corrective feedback and simplified probing rather than simply supplying answers. This structure encourages learners to monitor their own understanding and identify gaps in their knowledge, and thus helps to support the development of essential self-regulation skills.  This mode can be incorporated into a course by instructors for regular use, for example to facilitate end-of-week review sessions.

\textbf{Contextual and Authentic Learning}
By design, the `Code Query' feature of CodeHelp supports student learning in the context of authentic programming tasks. Students use the tool while working on real assignments, often outside class hours. Usage data from a semester-long deployment in an introductory computer and data science course (n = 52) showed sustained engagement over 12 weeks, with more than 2,500 student queries submitted \cite{sheese2024patterns}. Most queries focused on debugging and implementation questions that were directly tied to the course assignments, demonstrating highly contextual use.  The study also found a modest positive correlation between overall tool usage and course performance, suggesting that engagement with the tool did not undermine learning and may have supported it.

Students typically use the `Code Query' feature of CodeHelp after encountering a difficulty or error when programming, such as an error message or unexpected behaviour of their program.  This prompts them to reflect on what might be going wrong. CodeHelp responds with conceptual explanations and high-level guidance, after which students return to their code to revise and test their approach. This iterative pattern of attempt, reflection, refinement, and re-experimentation mirrors Kolb's experiential learning cycle (see Section \ref{sec:contextual_learning}) and situates learning within active problem solving.

\subsubsection{Adaptive Foundations of CodeHelp}

Although CodeHelp demonstrates strong alignment with the pedagogical foundations detailed in Section \ref{sec:learn-with-ai}, its adaptive capabilities are currently limited.  Nevertheless, it illustrates how pedagogically grounded systems can provide useful learning support even before full adaptivity is implemented.

As outlined in Section \ref{sec:capture}, the `Capture' stage of the adaptive cycle describes the importance of systematically recording digital traces in order to enable learner modelling and provide adaptive support.  The CodeHelp system does capture detailed interaction data, including all queries and responses, which can be reviewed and exported by instructors.  These conversational exchanges between the student and the AI model form digital footprints of learners' problem-solving processes.  Analyses reveal substantial variation across students in the kinds of help they seek, and their patterns of engagement \cite{sheese2024patterns}. This variation is particularly valuable for adaptive systems as it provides a richer signal for modelling.  Translating these captured traces into learner models that inform more dynamic adaptive support across sessions remains an important direction for future work.  

The adaptation that does occur currently in CodeHelp is limited to within individual sessions. For example, in the `Focused Tutors' mode, the system adjusts its responses and the probing questions that is asks based on how students answer. However, this adaptation does not persist across sessions, and the system does not automatically refine its behaviour based on accumulated data. Instructors may manually adjust learning outcomes or prompts in response to observed patterns, but this currently remains human-mediated rather than automatic. 

\subsubsection{Responsible Design Foundations of CodeHelp}

\textbf{Security and Privacy} The architectural design of CodeHelp provides support for important aspects of the Responsible Design pillar of the framework.  As an open-source system that can be self-hosted, institutions have the option to retain control over data storage and processing. The platform integrates with LMS environments via LTI, aligning with established institutional security practices. Importantly, even the hosted version of CodeHelp allows instructors to configure the AI backend using any OpenAI-compatible REST endpoint.  This design means that instructors and institutions have the option to use their own managed or locally hosted AI models.  However, responsibility for secure configuration ultimately rests with deploying institutions.

\textbf{Transparency and Inclusion} CodeHelp promotes transparency through the public availability of its source code and prompting pipeline, which means that the guardrail mechanisms can be reviewed. Student-facing responses include a notice that acknowledges outputs may not always be correct, highlighting the limitations of LLM-based systems.  One of the main motivations for the CodeHelp project was to lower barriers to help-seeking by offering private, always-available assistance, which can support inclusion given that in-person support often fails to reach all students equally \cite{smith2017digital}.  Research suggests that some learners do feel more comfortable interacting with the system than approaching instructors or TAs directly \cite{liffiton2023codehelp}. 

\textbf{Accountability} Instructor oversight is a core part of the system design. Dashboards allow instructors to monitor usage patterns and review and export queries and responses. With the `Focused Tutors' mode, instructors retain control over the suggested learning outcomes that CodeHelp generates from provided course material, and they can edit them to align with their pedagogical goals (see Figure \ref{fig:CodeHelpExample2}(b)). To further customise model responses, instructors can define an `avoid set' which is a set of keywords that should be avoided in any AI response to the student.  In addition, after each response the system asks students to indicate whether the response was helpful (see Figure~\ref{fig:CodeHelpExample2}(a)), creating an auditable feedback signal that can support review and ongoing refinement.  While the actual interactions are generated dynamically by the underlying language model, these mechanisms help to support instructor oversight of their learners' interactions.

\subsection{JeepyTA: An AI-Powered Teaching Assistant for Course Discussion and Formative Feedback} \label{sec:jeepyta} 

\subsubsection{Overview of the JeepyTA}

JeepyTA is an AI-driven teaching assistant developed by the Penn Center for Learning Analytics at the University of Pennsylvania, first launched in Fall 2023. The system employs a multi-turn conversational architecture powered by large language models and is model-agnostic, capable of operating with various LLMs including GPT, Llama, and DeepSeek. Deployments have utilized successive OpenAI GPT models, from GPT-3.5 Turbo through GPT-5.

As of late 2025, JeepyTA has been deployed across nearly 20 courses at four institutions in the United States and Singapore. Integrated into the open-source Flarum discussion forum where it appears as a distinctly marked AI Teaching Assistant, it can also be integrated into discussion forums in other platforms such as Canvas. A Progressive Web App architecture provides mobile accessibility with email and push notifications.

JeepyTA's capabilities span multiple pedagogical functions: answering logistical questions, participating in discussions, delivering targeted feedback on essays, projects, and code, and supporting brainstorming. To ensure alignment with specific course contexts, instructors prime JeepyTA with chosen reference materials: syllabi, textbooks, readings, and past feedback examples. These resources are embedded through a retrieval-augmented generation (RAG) workflow. Instructors can customize behavior by task, determining which areas JeepyTA addresses and the required oversight level, including options for human-in-the-loop review before responses become visible to students.

\begin{figure}[h]
    \centering
    \includegraphics[width=0.8\textwidth]{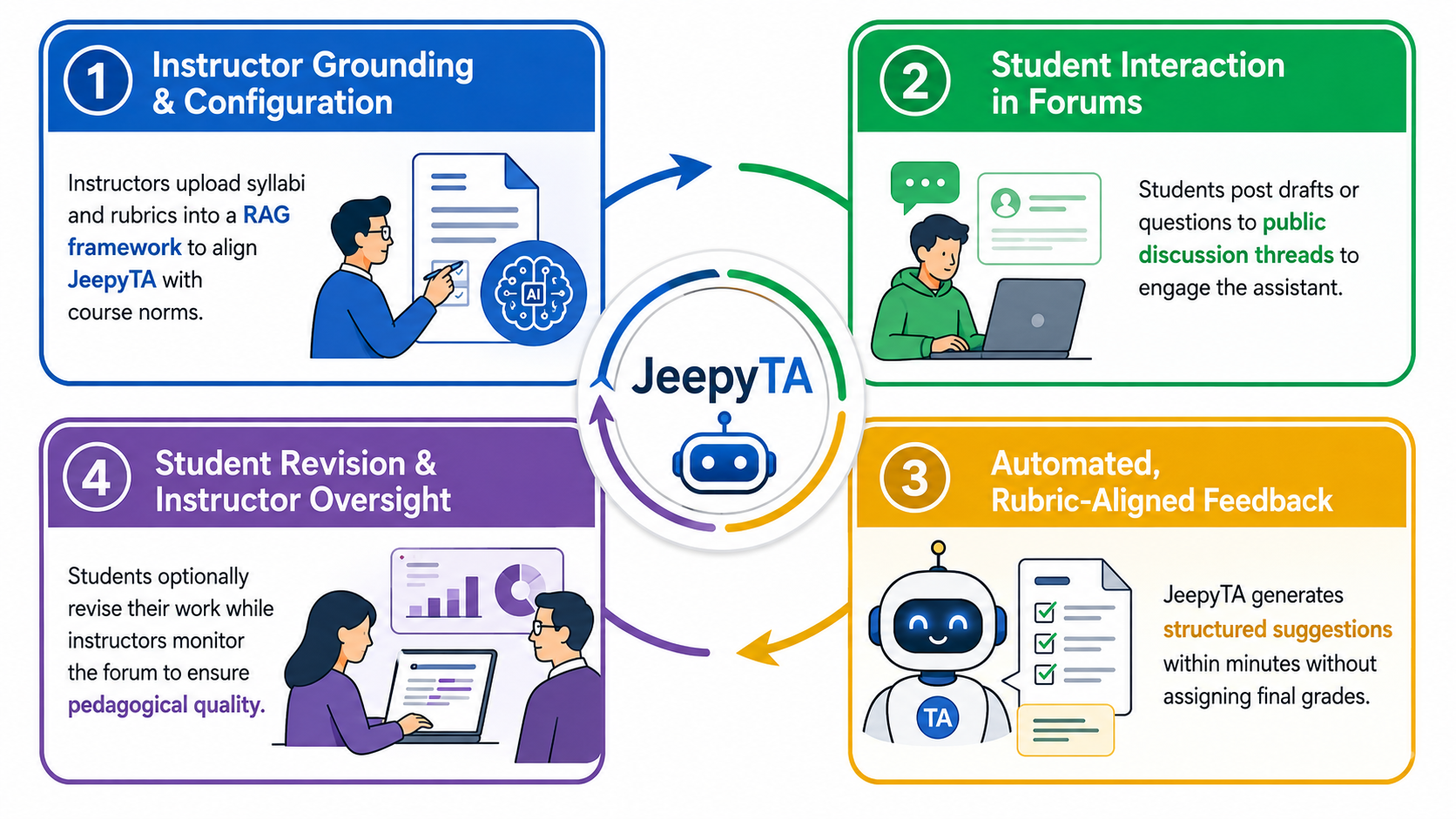}
    \caption{Overview of the JeepyTA workflow}
    \label{fig:JeepyTA}
\end{figure}

Figure~\ref{fig:JeepyTA} depicts JeepyTA's core workflow. The instructor selects a set of pedagogical functions for a specific course. When the student interacts with the relevant activities -- posting a question to the discussion forum or submitting an assignment for feedback -- JeepyTA generates a response. In some cases, the instructor reviews that feedback; in other cases the student receives it immediately. Next, the student chooses how to act based on the response, perhaps asking a follow-up question, submitting a revised assignment, or ending the discussion thread. This sequence of events happens multiple times per student, per course. 

\subsubsection{Pedagogical Foundations of JeepyTA}

\textbf{Deep and Interactive Learning.} JeepyTA engages students through sustained dialogue. When responding to reflections on readings and lectures, it acknowledges contributions, reinforces key ideas, and connects insights to course themes. It clarifies concepts by summarizing arguments and citing specific readings. When students introduce interesting perspectives, JeepyTA poses follow-up questions to encourage further discussion, supporting the iterative meaning-making process central to deep learning.

In brainstorming mode, JeepyTA serves as a collaborative partner. In a Games and Learning course, students consulted it to propose educational uses for video games through structured play journal assignments, with JeepyTA scaffolding connections between gameplay experience and pedagogical application.

\textbf{Guided Scaffolding.} When delivering feedback on student essays and projects, JeepyTA evaluates both higher-level conceptual elements—such as whether students have employed course theories in their arguments, detailed solution limitations, and appropriately addressed stakeholder needs—and lower-level aspects including argument clarity, evidence use, structure, and writing quality. In its prompt, JeepyTA is asked to ``provide actionable insights rather than shallow suggestions,'' offering concrete guidance.

For programming, JeepyTA identifies errors and suggests corrections without providing complete solutions, helping students develop diagnostic skills and problem-solving capacity. When students describe issues vaguely, JeepyTA asks clarifying questions about error messages, intended functionality, or steps already attempted, scaffolding the troubleshooting process itself.

The system's scaffolding approach incorporates instructor oversight mechanisms that balance immediacy with accuracy. Instructors can configure whether JeepyTA's responses appear immediately or await human review, adjustable by discussion category. This flexibility allows graduated release of autonomy. In practice, instructors often apply tighter control for conceptual questions, while enabling immediate formative feedback for assignments to support rapid iteration.

\textbf{Learning to Learn and Higher Order Learning.} JeepyTA supports metacognitive development through feedback and reflective prompts. When providing essay feedback, the system comments on strategic aspects of writing such as how arguments are constructed and how evidence is marshalled, making these higher-order processes visible and discussable. In debugging support, JeepyTA suggests higher-order strategies such as adding diagnostic print statements, checking variable values systematically, or decomposing complex functions into testable components.

However, the system remains primarily responsive rather than proactive in cultivating metacognitive awareness. While JeepyTA explains reasoning after student attempts, it rarely prompts students to articulate their own learning goals, monitor progress, or reflect on the evolution of their understanding. The weekly discussion summaries it offers within some courses (again at instructor discretion) represent one mechanism for prompting reflection, by highlighting key themes, recurring arguments, and unresolved questions. However, these summaries serve instructors and the collective more than individual metacognitive development.

\textbf{Contextual Learning.} JeepyTA situates learning authentically through course-specific grounding. In a course on Educational Data Mining, for example, JeepyTA was explicitly instructed to prioritize methodologically appropriate techniques for the domain rather than generic approaches that might be acceptable elsewhere.

Another mechanism used by JeepyTA for contextual learning is its persona functionality. In a course on cultural foundations for teaching and learning, JeepyTA adopted personas based on lived experiences of individuals from historically underrepresented groups, defined through description (name, role, context, cultural background), situation (interaction role, task, participants), and instruction (tone, detail, specific elements). By engaging with these personas—such as ``Felipe,'' a Mexican American teacher educator, or ``Claire,'' a Hmong American educator—preservice teachers encountered concrete examples of culturally relevant pedagogy, including specific strategies for incorporating funds of knowledge into classroom settings.

\subsubsection{Adaptive Foundations of JeepyTA}

JeepyTA's current implementation represents an early stage in realizing Section 4's adaptive foundations, with stronger development in some dimensions than others.

\textbf{Capture.} The system logs comprehensive conversational data: student queries, JeepyTA responses, instructor modifications, and follow-ups. The RAG architecture tracks which materials are retrieved per query, providing implicit evidence of which concepts students are engaging with or struggling to understand. Additionally, when JeepyTA provides essay feedback or debugging support, the student work itself can reveal conceptual understanding, strategic approaches, and error patterns. Recent analyses have leveraged these data to study how students adapt work following feedback.

However, capture remains limited compared to Section 4.1's vision. The system does not currently integrate data from beyond the forum, such as performance or time-on-task in other course components, or multimodal data.

\textbf{Model.} Learner modeling capability is minimal. JeepyTA maintains no persistent representations of individual students' knowledge, misconceptions, strategies, or affect. Each discussion thread is treated independently and is not supplemented by a cumulative student model.

This represents a significant gap relative to this paper’s adaptive foundation vision. The LLM's conversational memory within a specific discussion thread provides some continuity, but no other measurement of knowledge, affect, metacognition, or other key quantities is currently present.

\textbf{Adapt.} JeepyTA's adaptive behavior occurs at the level of RAG content retrieval and situational prompt design rather than deep personalization. The system can be configured with different behavioral rules by discussion category, but this is an instructor-level design choice rather than true system adaptivity.

\textbf{Evolve.} JeepyTA incorporates mechanisms for system-level improvement, but these are human-mediated rather than automated. Instructors review, approve, edit, or discard responses, with the option to remember these corrections for future interactions. This creates feedback loops where the knowledge base and behaviors improve based on instructor input. When instructors note recurring issues, such as JeepyTA defaulting to inappropriate problem-solving approaches or missing course-specific conventions, prompts can be refined to address these patterns. For example, instructors have iterated prompts for feedback and brainstorming based on prior semesters' experiences, but this remains in the realm of learning engineering rather than adaptive learning.

What remains absent is the kind of automated, continuous evolution described in Section 4.4, where the system employs A/B testing, multi-armed bandit algorithms, or reinforcement learning to discover which pedagogical strategies work best for which students in which circumstances.

\subsubsection{Responsible Design Foundations in JeepyTA}

JeepyTA's implementation demonstrates attention to several dimensions of responsible AI design, with particular emphasis on transparency and human oversight.

\textbf{Transparency.} JeepyTA implements several transparency mechanisms that align with the principles outlined in Section 5.2. Most fundamentally, the system is distinctly marked as an AI Teaching Assistant within the forum, ensuring students understand they are interacting with an automated system rather than a human TA or instructor. The system also provides transparent explanations in many interactions. JeepyTA often cites course materials explicitly. This allows students to verify information against original sources and understand the evidential basis for responses. When providing essay feedback, JeepyTA grounds its comments in rubric criteria, clarifying which standards are being applied and why particular aspects receive attention.

However, transparency is incomplete. It does not maintain open learner models that would make its interpretations visible for inspection and negotiation. Like many LLM-based systems, the underlying architecture remains a ``black box''—neither students nor instructors can trace how inputs and retrieved materials transform into outputs.

\textbf{Accountability.} JeepyTA is designed with a commitment to human oversight. The configurable review mechanism—allowing instructors to determine whether responses require approval before posting, adjustable by category—preserves instructor authority while enabling rapid feedback where appropriate. JeepyTA handles high-volume, routine queries and provides swift formative guidance; human educators remain responsible for conceptual instruction and high-stakes decisions. The system preserves audit trails of all interactions, supporting retrospective analysis, continuous improvement, and accountability when issues arise.

\textbf{Inclusivity.} JeepyTA incorporates several inclusivity-oriented features, though this remains an area for further development. Its 24/7 availability supports equitable access for students with non-traditional schedules, caregiving responsibilities, or different time zones. Students may also receive faster responses from human instructors and TAs, who can focus attention on situations requiring human expertise (Liu et al., in press). JeepyTA additionally interacts in multiple languages, leveraging the multilingual capabilities of its underlying LLMs.

Beyond this, the persona functionality explicitly centers diverse cultural perspectives and funds of knowledge. By grounding personas in lived experiences of individuals from historically underrepresented groups, the system exposes learners to a wider range of worldviews. This approach is based on the premise that inclusive AI should not only avoid bias but also actively support engagement with diverse ways of knowing and being.

\subsection{Recast: A university platform for AI Companions} \label{sec:recast}

\subsubsection{Overview of Recast}
The University of Technology Sydney (UTS) is a public institution serving approximately 50,000 students studying in hybrid and fully online modes, in Sydney and abroad. UTS has developed a platform called \textit{Recast} to support the design and publishing of AI Companions, (termed Recast Assistants) for staff and students. Recast provides an Assistant design interface enabling staff to design and test system prompts, adjust the usual dialogue parameters such as temperature and strictness, and optionally ground the Assistant in a folder of documents for Retrieval Augmented Generation (RAG) if it is desired that its answers are sourced only from that curated corpus (hence, Assistants do not perform dynamic internet searches: Microsoft Copilot is provided for this). Role-based permissions manage student and staff access to Assistants via their university account, making this a formal part of the university’s learning technology ecosystem that can be used for formal assessments. Assistants can be set to offer voice input/output, while custom prompting reflects the specific pedagogy, topics and learning outcomes of their course, or a specific assignment for which the Assistant was designed. 

\subsubsection{Pedagogical foundations of Recast}
Since Recast is a platform for designing and deploying diverse types of Assistant, there is no single tool to describe. Rather, families of Assistant are emerging, with correspondingly diverse pedagogical bases, conducting distinct kinds of dialogue. Some of these are introduced next, mapped to two of the processes in the pedagogical foundation introduced above. As can be seen, these are focused less on coaching mastery of curriculum concepts, and more on higher order, contextualised skills.

\textbf{Learning to learn \& higher order learning.} \textit{Reflection Assistants} conduct a strongly scaffolded student reflection on some aspect of their course. An example would be reflection on the progress students have made on goals they set for themselves at the start of semester, prior to submitting a written reflection. These submitted reflections can then be analysed using an LLM for evidence of students’ affective engagement and belonging \cite{RN13362}. \textit{ Critical Thinking Assistants} use a form of Socratic questioning about the implicit premises behind students’ questions, to help students check their assumptions and refine their research questions \cite{RN13587}. This has been used in courses including Data Science and Sustainable Futures, as well as by PhD students and academics.

\textbf{Contextual learning.} \textit{Role-Play Assistants} provide authentic, experiential learning through voice-based role-play simulations \cite{RN13587}. These enable students to practice conducting challenging conversations within the dynamics of a verbal interaction (in contrast to slower, written dialogue). At the time of writing, we have such Assistants integrated into Health and Law assessments, whose roles and conversations are, as one would expect, very different. Snapshots of these and other Assistants are provided online,\footnote{UTS:CIC GenAI Apps \& Analytics: https://cic.uts.edu.au/tools/genai-apps-and-analytics} and empirical studies are now underway.

\subsubsection{Adaptive foundations of Recast}
Recast Assistants are not adaptive in the sense that an ITS continually updates a persistent learner model with the student’s level of competence mapped to a detailed curriculum model of requisite knowledge and skills. However, within the capability of an LLM with custom prompting and grounding, the Assistants \textit{capture }all interactions persistently (opening the possibility of running advanced analytics on the transcripts), and \textit{adapt} in the sense that their responses are contingent on what a student says, guided and constrained by diverse forms of pedagogically-tuned system prompt. 

\subsubsection{Responsible design foundations of Recast}
In conceiving, designing, implementing and evaluating Recast as a trustworthy institutional platform for AI Companions, it has become evident that responsible design is a complex, sociotechnical process, interweaving technical, user experience, policy and governance practices. The following sections explain the choices UTS has made plus rationale. However, this remains a turbulent landscape: as UTS gains governance experience in deploying Recast at scale, as empirical evidence builds on student and educator experiences, and as technological capabilities and limitations become apparent, these design decisions will be under regular review.

\textbf{Security and privacy.} Recast had to pass rigorous cybersecurity and architecture audits to be part of the UTS enterprise computing infrastructure. LLMs are hosted securely within an enterprise AI vendor with all data storage within state, while user authentication is handled by the institutional identity and permissions layers, drawing from the enrolment database to control which Assistants a student sees. All Assistant transcripts are archived, since UTS considers itself accountable for how its platforms are being used, must be able to verify performance for quality assurance, and investigate any incidents. In institutional ethics-approved research projects, student identifiers are not included in transcripts and replaced with new identifiers to enable dataset merging.

\textbf{Transparency on transcript privacy.} A particular context for transparency concerns the privacy of Recast transcripts. Unlike Khanmigo, our students are adults, and the choice was made that the teaching team cannot by default see student transcripts. Students are informed that all Recast sessions are logged, but that their individually identifiable chats are not being monitored by academics/tutors. The rationale for this is it would likely disincentivise student adoption, or impair engagement, given the evidence that students value chatbots as private support, avoiding fear of embarrassment or trepidation in approaching staff \cite{RN13435}. A teacher dashboard provides cohort-level quantitative indicators of engagement, which in the future could include more advanced, research-informed transcript analytics tuned to the nature of the dialogue. There are two exceptions: (i) the lead academic can view deidentified transcripts if students have given informed consent to participate in an ethics-approved research project; (ii) tutors can view identifiable transcripts when the Recast session is a formal requirement for an assignment in which the transcript is a primary artifact for assessment (e.g., a role-play dialogue).

\textbf{Accountability.} Firstly, there is accountability for AI ethics. In 2021, we initiated a set of student and staff “EdTech Ethics” consultations, which proved to be a rewarding demonstration of the value that a deliberative democracy process can bring to organisation-wide deliberation \cite{RN13155}. Subsequently, UTS formed a cross-university AI Operations Board, the governance body that implements our AI Operations Policy, which in turn builds on the state’s AI ethics framework. Proposals for Recast Assistants are governed by the same review process that covers all new AI applications within UTS. This introduces a second form of accountability, that we have confidence in an Assistant’s trustworthiness. Recast quality assurance requires developmental testing by the teaching team, plus automated testing that exhaustively verifies appropriate responses to a range of likely prompts, wellbeing alerts, as well as jailbreaking efforts. Finally, there is accountability for student wellbeing. Content filtering is designed to identify concerning material from either the student or LLM, at which point Recast declines to engage in further discussion. If student distress is detected, guardrails in the system prompt link students to UTS Counselling. Filters can be adjusted at the individual Assistant level, so that students can legitimately discuss sensitive topics for their studies (e.g., in Forensics, or Social Work).

\textbf{Inclusivity.} It was a requirement that the user interface meet W3C accessibility standards, which underwent testing by the UTS accessibility specialist team. However, there are degrees of accessibility, and there remain aspects that should be improved.

To summarise, the Recast case demonstrates that in order to implement the responsible design framework principles in an AI Companion platform operating in the enterprise IT at institutional scale, technical choices under the control of developers alone are critical, but insufficient: they must also align with transparent communication by the teaching team to students about who can see what, and with the institution’s policy and governance mechanisms. There is no context-independent blueprint for how to implement AI companions responsibly, but this case demonstrates how and why one institution made its choices.

\begin{figure}[t]
    \centering
    \includegraphics[width=0.95\textwidth]{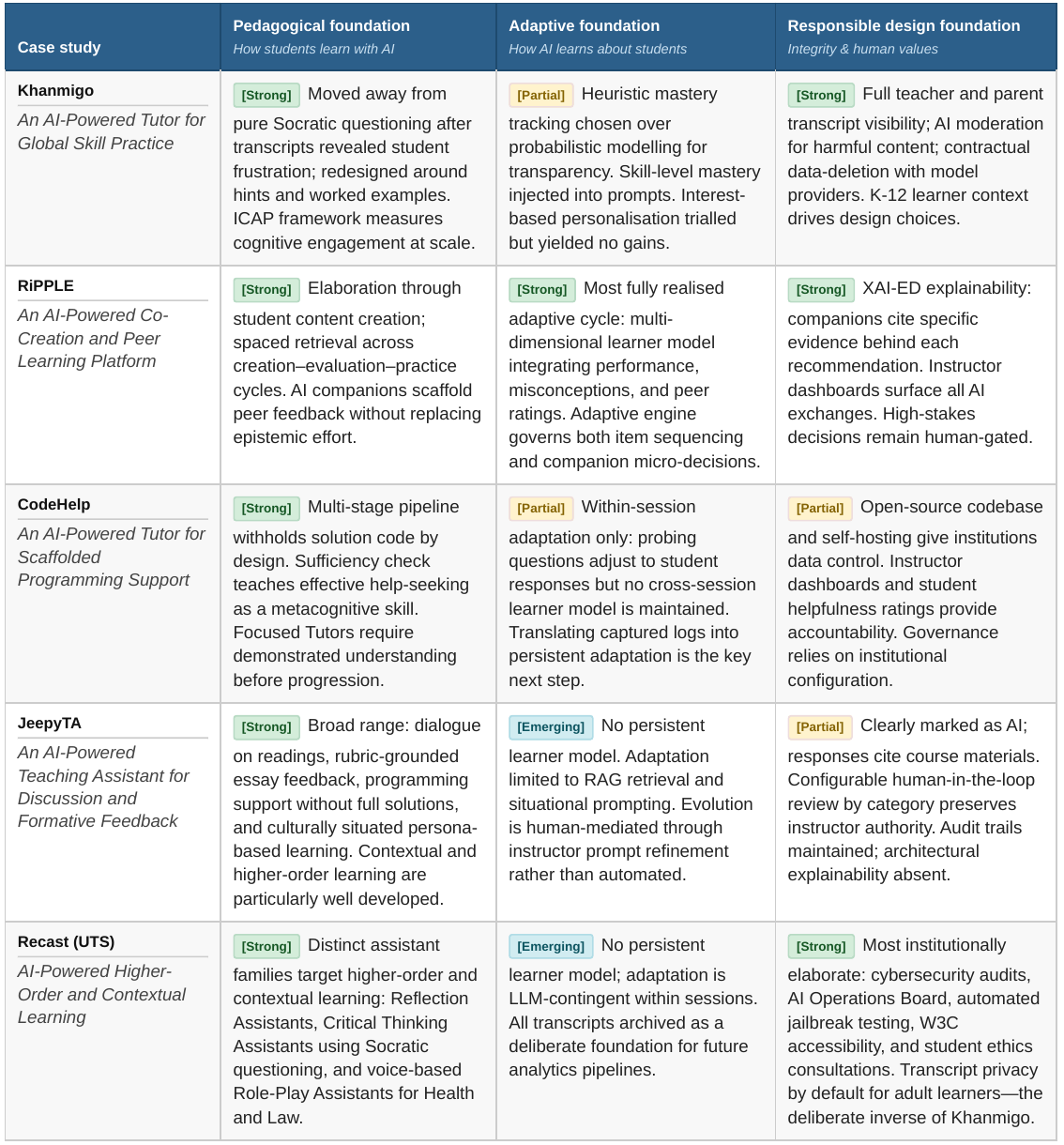}
    \caption{Synthesis of case studies mapped against the three framework pillars.}
    \label{tab:synthesis}
\end{figure}

\subsection*{Pedagogical, Adaptive, and Responsible Foundations in Practice}

Three patterns emerge from the case studies, aligned with the framework (see Table~\ref{tab:synthesis}).  First, the pedagogical foundation is the most mature across all cases, but different key mechanisms are used by the tools. Khanmigo and CodeHelp focus on avoiding the generation of direct solutions in order to preserve productive struggle; RiPPLE shifts learners into generative roles as authors, peer reviewers, and active practitioners; JeepyTA embeds support across a range of activities that are highly course-specific and contextualised; and Recast extends the idea of a learning companion beyond content and into reflection, critical thinking, and authentic role-play. Across these cases, a central pedagogical design decision is around which cognitive work should remain with the learner.

Second, the adaptive foundation reveals the greatest variation in maturity. RiPPLE most closely approximates the full adaptive cycle, with persistent learner modelling and personalised recommendation mechanisms already embedded in the platform. Khanmigo uses a narrower form of adaptivity by incorporating mastery information about the current and prerequisite skills into tutoring interactions. CodeHelp, JeepyTA, and Recast all capture some traces of learner interaction and adapt within sessions or contexts, but the way they adapt remains largely human-mediated rather than automatic via learner models. Persistent, longitudinal adaptivity remains the most significant gap between current implementations and the full vision of an AI learning companion.

Third, the responsible design foundation is highly context-dependent. Khanmigo's full conversation visibility is appropriate to a school-based context, whereas Recast's transcript privacy by default reflects a university context in which adult learners may need private space to experiment, and ask questions that may reveal a lack of understanding or uncertainty. CodeHelp and JeepyTA provide for instructor configurability and auditability, while RiPPLE emphasises explainable learner modelling and instructor oversight of learner-generated content. These differences show that responsibility depends very much on the learner and institutional contexts in which the companion operates.

The five case studies demonstrate that current AI-powered educational tools are beginning to progress beyond generic task-completion tools, although progress with respect to the foundations of our AI companion framework is uneven. Their strongest common contribution lies in pedagogical design where each tool attempts, in different ways, to keep learners cognitively active and reflective. One challenge ahead is to close the adaptive loop, and using persistent evidence about learners to personalise support without compromising privacy, agency, or human oversight.  Ultimately, learning-oriented AI should be judged by how deliberately it supports the processes through which durable learning develops.

\section{Conclusion}
This paper has argued that the growing integration of LLMs into education demands more than better prompting or pedagogical guardrails applied to tools built for other purposes. It requires a fundamental reconceptualisation of what AI in education is for. The learning-performance paradox reveals that AI optimised for task completion can actively undermine the processes through which durable learning occurs. The response we have proposed is a distinct class of AI agents, AI learning companions, designed to be embedded in educational environments with an explicit mission to prioritise learning over performance. The framework we have presented organises companion design around three interrelated foundations: a pedagogical foundation concerned with how students learn with AI, an adaptive foundation concerned with how AI learns about students, and a responsible design foundation concerned with how companions act with integrity and uphold human values.

The five case studies, each authored by researchers who led or made significant contributions to the tools described, provide firsthand insight into how these foundations play out in practice across diverse educational contexts, levels, and design approaches. They collectively demonstrate that deliberate pedagogical design can meaningfully distinguish learning-oriented AI from generic task completion tools, and that responsible design looks different depending on whether learners are minors or adults, whether tools are purpose-built or platform-based, and whether the goal is content mastery or higher-order development. Persistent adaptivity remains largely emerging across most cases, pointing to the most important directions for future work: how persistent learner modelling can be developed without compromising privacy or learner agency; how delayed retention and transfer can be built into evaluation frameworks as standard practice rather than an afterthought; and how governance models can be adapted and scaled across diverse institutional and regulatory contexts.

Ultimately, the promise of AI learning companions lies not in replacing human teaching or accelerating task performance, but in cultivating learners who are more reflective, more metacognitively aware, and better equipped to learn independently in an AI-rich world. The shift from optimising for performance to cultivating learning is not a minor adjustment to how we deploy AI in education. It is a fundamental reorientation of what we ask AI to do for learners, and for education itself.

\bibliographystyle{abbrv}
\bibliography{sn-bibliography}

\end{document}